\documentclass[prb,twocolumn,showpacs,amsfonts,amssymb,floats,superscriptaddress,aps]{revtex4}

\usepackage[final,dvips]{epsfig}
\newcommand{\pd}{{\phantom{\dagger}}}

\begin{document}

\title[]
{Dissipative Two-Electron Transfer}
\author{Sabine Tornow} \affiliation{Theoretische Physik III, Elektronische Korrelationen
und Magnetismus, Institut f\"ur Physik, Universit\"at Augsburg,
86135 Augsburg, Germany} \affiliation{ Institut f\"ur Mathematische
Physik, TU Braunschweig, 38106 Braunschweig, Germany }
\author{Ralf Bulla}
\affiliation{Theoretische Physik III, Elektronische Korrelationen
und Magnetismus, Institut f\"ur Physik, Universit\"at Augsburg,
86135 Augsburg, Germany} \affiliation{Institut f\"ur Theoretische
Physik, Universit\"at zu K\"oln, 50937 K\"oln, Germany}
\author{Frithjof B. Anders}
\affiliation{Fachbereich Physik, Universit\"at Bremen, 28334 Bremen,
Germany}
\author{Abraham Nitzan }

\affiliation{School of Chemistry, The Sackler Faculty of Exact
Sciences, Tel Aviv University, Tel Aviv 69978, Israel}


\begin{abstract}

We investigate non-equilibrium two-electron transfer in a model
redox system represented by a two-site extended Hubbard model and
embedded in a dissipative environment. The influence of the
electron-electron interactions and the coupling to a dissipative
bosonic bath on the electron transfer is studied in different
temperature regimes. At high temperatures Marcus transfer rates are
evaluated and at low temperatures, we calculate equilibrium and
non-equilibrium population probabilities of the donor and acceptor
with the non-perturbative Numerical Renormalization Group approach.
We obtain the non-equilibrium dynamics of the system prepared in an
initial state of two electrons at the donor site and identify
conditions under which the electron transfer involves one concerted
two-electron step or two sequential single-electron steps. The rates
of the sequential transfer depend non-monotonically on the
difference between the inter-site and on-site Coulomb interaction
which become renormalized in the presence of the bosonic bath. If
this difference is much larger than the hopping matrix element, the
temperature as well as the reorganization energy, simultaneous
transfer of both electrons between donor and acceptor can be
observed.
\end{abstract}
\pacs{71.27.+a, 34.70.+e, 82.39.Jn } \maketitle

\section{Introduction}
Electron transfer is a key process in chemistry, physics and biology
\cite{Jortner,May,Marcus,Nitzan} encountered in, e.g., chemical
redox processes, charge transfer in semiconductors and the primary
steps of photosynthesis. In condensed polar environments the process
involves strong coupling to the underlying nuclear motion and is
usually dominated by the nuclear reorganization that accompanies the
charge rearrangement. A quantum mechanical description of electron
transfer in such a dissipative environment is given by the
spin-boson model \cite{Leggett,Weiss} and its variants; this model
accounts for the essential energetics and dynamics of the process,
such as the non-monotonic dependence of the transfer rate on the
energy asymmetry, the energy difference between the initial and
final electronic states.

Although standard descriptions of such processes focus on
single-electron transfer \cite{Nitzan,Jortner,Leggett,Weiss},
two-electron transfer has been suggested as the dominant mechanism
in some bioenergetic processes that occur in proteins
\cite{proteins,MayPRE}, transfer in transition metal complexes
\cite{ru,fe}, electrode reactions \cite{Gileadi}, artificial
photosynthesis and photoinduced energy- and electron-transfer
processes \cite{keene}, biological electron transfer chains
\cite{nature}, transfer in fuel cells \cite{fuel} as well as in DNA
\cite{DNA}. Further examples are selfexchange reactions such as
Tl(I)/Tl(III) and Pt(II)/Pt(IV) \cite{Zusman2} and electron-pair
tunneling \cite{vonoppen,alex,Chakra} in molecular electronic
devices.

The theoretical description of two-electron transfer dynamics
differs fundamentally from its single-electron counterpart. More
than two states have to be considered \cite{Zusman,Okada} and
electron correlations induced by the Coulomb repulsion and the
coupling to the environment need to be accounted for. Usually, the
on-site Coulomb interaction in molecules is much larger than the
inter-site interaction. \cite{Fulde,Campbell,Starikov} However, due
to the polarization of the local environment the short range
interaction may be strongly screened. Then, the inter-site
interaction $V$ can be of the same order or even exceed the on-site
Coulomb interaction $U$. \cite{Brink,Starikov} While $U$ favors a
homogeneous charge distribution, the inter-site interaction $V$
inclines spatially inhomogeneous charge accumulation. Since the
non-equilibrium dynamics is governed by the energy difference $U-V$,
the competition between both interactions influences strongly the
type of charge transfer dynamics. Depending on the sign of the
energy difference a single concerted two-electron step or two
sequential single-electron steps may occur.

In this paper we consider a system comprised of a donor (D) and an
acceptor (A) site. They share two electrons which are coupled to a
non-interacting bosonic bath. Such a donor-acceptor system has four
different states: two doubly occupied donor ($D^{2-}A$) and acceptor
($DA^{2-}$) states, and two degenerate states $D^-A^-$ with one
electron each on the donor and acceptor site (with different spin).
Their energy difference depends on the difference between on-site
and inter-site Coulomb repulsion as well as the bias $\varepsilon$
which we do not consider here. The transition $ D^{2-}A \rightarrow
DA^{2-} $ occurs as a concerted transfer of two electrons or an
uncorrelated sequence of one-electron transfer events during which
the intermediate $D^- A^-$ is formed. The transfer rate of each
electron may be different and shows a non-monotonic behavior on the
energy asymmetry between the states. In this paper, we are mapping
conditions under which the system performs concerted two-electron
transfer or a sequential single-electron process. To this end we
study the non-equilibrium dynamics of the donor-acceptor system
initially prepared with two electrons at the donor site. We evaluate
the rates for single-electron transitions and an electron-pair
transfer in different regimes of the Coulomb repulsion and
environmental response.

The occurrence of such a correlated electron-pair transfer can be
already understood within a donor-acceptor system, decoupled from
the environment in which the strong on-site Coulomb repulsion
exceeding considerably  the inter-site repulsion. We start from a
doubly occupied excited donor state $D^{2-}A$ compared to the
$D^-A^-$ ground states. Energy conservation implies a concerted
electron transfer. If the transfer matrix element $\Delta$ is much
smaller than this energy difference the transfer occurs as a
tunneling process of an electron pair in which the intermediate
states $D^-A^-$ are occupied only virtually analogous to a
"superexchange" process (see, e.g., Ref.~\onlinecite{Jortner}).

In the present paper we investigate the effect of coupling to a
dissipative bosonic environment, with a total number of two
electrons occupying donor and acceptor sites. These two electrons
experience the on-site Coulomb repulsion $U$ when occupying the same
site and the Coulomb repulsion $V$ when occupying different sites.
In this paper we restrict ourselves to the simplest case where donor
and acceptor are each modeled by a single molecular orbital. In such
a system the difference $\tilde{U}=U-V$ is crucial for the dynamics.
The coupling to the bosonic bath has two major effects: (i) the
renormalization \cite{Mahan,Hewson} of the on-site Coulomb repulsion
$\tilde{U}$ to $\tilde{U}_{\rm eff}$ and (ii) dephasing as well as
dissipation of the energy from the donor-acceptor system to the
bath. The latter leads to the damping of coherent oscillations that
would otherwise exist between the quantum states of the related
molecule and, beyond a characteristic coupling strength, to
incoherent dynamics of the electron transfer process. These
considerations lead us to a dissipative two-site Hubbard model, a
minimal model that captures the essential physics comprising
correlations between electrons and their coupling to the dissipative
environment. It is discussed in detail in Sec.~\ref{model}. For a
comparison to experimental results it has to be supplemented by {\it
ab initio} calculations of the parameters.

The equilibrium properties of the model have been previously
studied\cite{TTB1} using the numerical renormalization group (NRG),
and  the real-time dynamics has been investigated\cite{MAK1,MA1}
using a Monte-Carlo technique at high temperatures where only
incoherent transfer is present.  In these Monte-Carlo calculations,
the effective Coulomb interaction was chosen to be $\tilde{U}_{\rm
eff}>0$ and no electron-pair transfer has been reported.
Two-electron transfer in a classical bath has been discussed in
Ref.~[\onlinecite{Zusman}] in the framework of three parabolic
potential surfaces (for the four states $D^{2-}A$, $D^-A^-$ and
$DA^{2-}$) as a function of a single reaction coordinate. A
generalization to donor-bridge acceptor systems is given in Ref.
[\onlinecite{MayPRE,MayJCP}].

Although two-electron transfer was observed in some regimes of
system parameters in the high-temperature limit, considering a
classical bath, it seems reasonable to expect that, at least between
identical centers, electron-pair tunneling processes are
particularly important at temperatures corresponding to energies
smaller than the effective energy difference between initial and
intermediate states $\tilde{U}_{\rm eff}$. At these temperatures
single-electron transfer cannot be activated (see Section
\ref{nonequi}). Therefore, we focus on the low-temperature regime
where the transfer is dominated by nuclear tunneling and where the
bosonic bath has to be treated quantum mechanically. Due to the
nuclear tunneling the electron transfer rate is constant over a wide
temperature range from zero temperature up to temperatures where
thermal activation becomes more important \cite{deVault}. In this
low-temperature regime, we employ the time-dependent
NRG\cite{Costi,AS1,AS2} (TD-NRG) which covers the whole parameter
space from weak to strong dissipation. The NRG is an accurate
approach to calculate thermodynamics and dynamical properties of
quantum impurity models.\cite{Wil75,Kri80,BLTV,BTV} For further
details of the NRG we refer to the recent review\cite{rev} on this
method.

The paper is organized as follows. In Sec.~\ref{model} we introduce
the model. Its high temperature behavior obtained from the Marcus
theory is described in  Sec.~\ref{section_toymodel}. Section
\ref{sec_NRG} introduces the NRG method, its extension to
non-equilibrium and its application to the present problem. In order
to gain a better understanding of the non-equilibrium dynamics
presented in Sec.~\ref{nonequi}, we summarize the equilibrium
properties of the model in the Sec.~\ref{equil}. We present a
detailed discussion of the real-time dynamics in Sec.~\ref{nonequi}.
Therein, we focus on the time evolution of occupation probabilities
of the different electronic states as the key observables. In
particular, when the dynamics can be described in terms of rate
processes, the dependence of the single and electron-pair rate on
the Coulomb repulsion parameters is analyzed. A summary of our
results is given in Sec.~\ref{sum}.

\section{model \label{model}}

We consider a model of a two-electron/two-site system coupled to a
bosonic bath. It is defined by the Hamiltonian

\begin{eqnarray}
  H=H_{\rm el} + H_{\rm coupl} + H_{\rm b} ,
\label{eq:ebm}
\end{eqnarray}
with
\begin{eqnarray}
H_{\rm el} &=& \sum_{\sigma,i={\rm A,D}} \varepsilon_i
c^\dagger_{i\sigma} c^\pd_{i\sigma}
     - \Delta \sum_\sigma \left( c^\dagger_{{\rm D}\sigma} c^\pd_{{\rm
A}\sigma} +                    c^\dagger_{{\rm A}\sigma} c^\pd_{{\rm
D}\sigma}                         \right)   \nonumber \\
    &+& U \sum_{i={\rm A,D}}  c^\dagger_{i\uparrow} c^\pd_{i\uparrow}
                              c^\dagger_{i\downarrow} c^\pd_{i\downarrow}
     +  \frac{V}{2} \sum_{\stackrel{\sigma,\sigma',i,j={\rm A,D}}{ ( i \neq j)}}  c^\dagger_{i\sigma} c^\pd_{i\sigma}
                              c^\dagger_{j\sigma'}
                              c^\pd_{j\sigma'} \ , \nonumber
\end{eqnarray}
\begin{eqnarray}
H_{\rm coupl} =
    \sum_{\sigma,i=A,D}\left(g_{\rm i} c^\dagger_{i \sigma} c^\pd_{i \sigma} \right)\sum_{n}
         \frac{\lambda_n}{2} \left(  b_{n}^{\dagger} + b^\pd_{n}
         \right), \nonumber
\end{eqnarray}
and
\begin{eqnarray}
H_{\rm b}=
         \sum_{n} \omega_{n} b_{n}^{\dagger} b^\pd_{n}
    \  ,
\nonumber
\end{eqnarray}

where $c_{i\sigma}$ and $c^{\dagger}_{i\sigma}$ denote annihilation
and creation operators for fermions with spin $\sigma$ on the donor
($i={\rm D}$) and acceptor ($i={\rm A}$) sites. The Hamiltonian
$H_{\rm el}$ corresponds to an extended two-site Hubbard model, with
on-site energies $\varepsilon_i$, hopping matrix element $\Delta$,
on-site Coulomb repulsion $U$ and an inter-site Coulomb repulsion
$V$ between one electron on the donor and one electron on the
acceptor. The difference $\tilde{U}=U-V$ measures the excess energy
needed to move an electron between the two sites. Such a two-site
Hubbard model without coupling to a bosonic bath has been
investigated in the context of electron transfer in
Ref.~\onlinecite{OR1}.

\begin{figure}
\vspace*{0cm} \epsfxsize=6cm \centerline{\epsffile{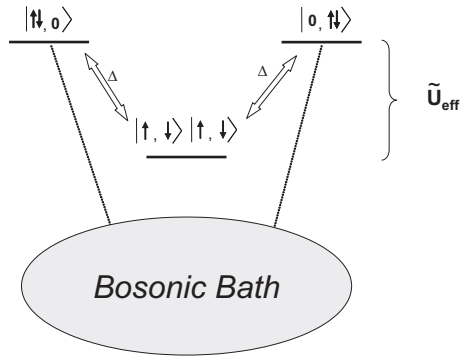}}
 \caption{The four states of model (eq. (1)) for the symmetric case ($\varepsilon=0$). The energy difference between
 the doubly occupied donor ($D^{2-}A$) or acceptor ($DA^{2-}$) and singly occupied donor acceptor pair ($D^-A^-$)
 depends on the effective renormalized interaction
 $\tilde{U}_{\rm eff}$ defined in eq. \ref{Ueff}.} \label{figstates}
\end{figure}

The Hamiltonian $H_{\rm b}$ models the free bosonic bath, with boson
creation and annihilation operators $b_{n}^{\dagger}$ and $b_{n}$,
respectively. The electron-boson coupling term, $H_{\rm coupl}$, has
the standard polaron form with the coupling constant for donor and
acceptor given by $g_{\rm D}\lambda_n$ and $g_{\rm A}\lambda_n$,
respectively. In what follows we set $\epsilon_{\rm
D}=-\epsilon_{\rm A}=\frac{\varepsilon}{2}$ and $g_{\rm A}=-g_{\rm
D}=1$. The latter choice implies that the polar bath is coupled to
the change in the electronic density $\sum_{\sigma} (c_{{\rm A}
\sigma}^{\dag} c_{{\rm A} \sigma}-c_{{\rm D}\sigma}^{\dag}c_{{\rm
D}\sigma})$:
\begin{eqnarray}
H_{\rm coupl}=\sum_{\sigma} (c_{{\rm A} \sigma}^{\dag} c_{{\rm A}
\sigma}-c_{{\rm D}\sigma}^{\dag}c_{{\rm D}\sigma}) \sum_{n}
         \frac{\lambda_n}{2} (  b_{n}^{\dagger} + b^\pd_{n}).
\end{eqnarray}
This two-site electron-boson Hamiltonian conserves the number of
electrons $\sum_{i \sigma} c_{i \sigma}^{\dagger} c_{i \sigma}$ and
the square of the total spin $\vec{S}^2$ as well as its
$z$-component $S_z$. The Hilbert space can therefore be divided into
different subspaces. In the subspace with one electron and
$S_z=1/2$, the model is equivalent to the spin-boson model
\cite{TTB1}. Here, we consider the subspace with two electrons and
$S_z=0$  which is spanned by the states
$|1\rangle=|\uparrow\downarrow,0 \rangle$, $|2\rangle=|
\downarrow,\uparrow \rangle$, $|3\rangle= |\uparrow, \downarrow
\rangle$, and $|4\rangle=|0,\uparrow\downarrow \rangle$ with the
notation $\vert A,D \rangle$ describing the occupation at the donor
($D$) and acceptor ($A$) sites. The four-dimensional basis in the
two-electron subspace is displayed in Fig. \ref{figstates}. We
define the following observables
\begin{eqnarray}
  \hat d_{D} &=& |1\rangle \langle 1|
\nonumber \\
  \hat d_{A} &=& |4 \rangle \langle 4|
\nonumber \\
\hat n_{AB} &=&  |2 \rangle \langle 2|+ |3 \rangle \langle 3|
\label{eqn:observables}
\end{eqnarray}
which measure the doubly occupancy $\hat d_{D}$ ($ \hat d_{A}$) on
the donor (acceptor) site and $\hat n_{DA}$ the combined population
of the states $|\uparrow,\downarrow\rangle$ and
$|\downarrow,\uparrow\rangle$. Note that in some works the states
$|\!\uparrow \downarrow,0\rangle$ and $|0,\downarrow \uparrow
\rangle$ are referred to as localized states \cite{MAK1}. We call
them doubly occupied states while the term localization is used
below for the self-trapping mechanism.

Consider the $4\times4$ Hamiltonian matrix in the electronic
subspace $( M)_{ij}=\langle i \vert H \vert j \rangle$
($i,j=1,\ldots, 4$). Introducing the notation
\begin{equation}
    \hat{Y} = \sum_{n} \omega_{n} b_{n}^{\dagger} b^\pd_{n}
    \ \ , \ \
    \hat{X} = \sum_{n}
         \lambda_n \left(  b_{n}^{\dagger} + b^\pd_{n}  \right)
    \ ,
\end{equation}
and shifting the Hamiltonian by a constant $V$ leads to

\begin{eqnarray} \left(
\begin{array}{cccc}
         \varepsilon +\tilde{U} + \hat{X} + \hat{Y} & -\Delta & -\Delta & 0\\
         -\Delta & \hat{Y} & 0 & -\Delta\\
         -\Delta & 0 &  \hat{Y} & -\Delta \\
         0 & -\Delta & -\Delta & -\varepsilon+   \tilde{U} - \hat{X} + \hat{Y}
 \end{array}     \right) \  ,
\end{eqnarray}
with $\tilde{U}=U-V$. Therefore, the dynamics of the system is
governed by the energy difference $\tilde{U}$ which replaces the
on-site Coulomb repulsion $U$. If screening of the local Coulomb
repulsion $U$ \cite{Brink,Starikov} is sufficiently large,
$\tilde{U}$ changes its sign and become effectively attractive. A
large inter-site  Coulomb repulsion $V$ favors an inhomogeneous
charge distribution.

It is convenient to rewrite the diagonal matrix elements of the
doubly occupied states in the form
\begin{eqnarray}
\langle 1|H|1\rangle=\varepsilon +\tilde{U}_{\rm eff}+\sum_n
\omega_n \left(b_n^{\dag} + \frac{\lambda_n}{\omega_n}\right)
\left(b_n + \frac{\lambda_n}{\omega_n}\right)  , \label{tomarcus1}
\end{eqnarray}
and
\begin{eqnarray}
\langle 4|H|4\rangle=-\varepsilon +\tilde{U}_{\rm eff}+ \sum_n
\omega_n \left(b_n^{\dag} - \frac{\lambda_n}{\omega_n}\right)
\left(b_n - \frac{\lambda_n}{\omega_n}\right) .  \label{tomarcus2}
\end{eqnarray}
Compared with the matrix elements of states corresponding to
$D^-A^-$
\begin{eqnarray}
\langle 2|H|2\rangle=\langle 3|H|3\rangle=\sum_n \omega_n b_n^{\dag}
b_n , \label{tomarcus3}
\end{eqnarray}
we can easily see that the electron-boson coupling generates an
effective renormalized interaction
\begin{eqnarray}
\tilde{U}_{\rm eff} =\tilde{U}- \sum_n \frac{\lambda_n^2}{\omega_n}
\ . \label{Ueff}
\end{eqnarray}
The renormalized interaction $\tilde{U}_{\rm eff}$ determines the
energy difference between $D^{2-}A$ ($DA^{2-}$) and $D^-A^-$, and
constitutes the only Coulomb interaction parameter in the present
model. The renormalization stems from a boson-induced effective
electron-electron interaction, already familiar from the Holstein
model \cite{Hewson}. Note that an artificial energy shift is present
in the single-electron subspace (spin-boson model)\cite{Weiss},
however the two states $|\uparrow\rangle$ and $|\downarrow \rangle$
are shifted in the same direction which can be handled by resetting
the zero of energy.

In analogy to the spin-boson model \cite{Leggett,Weiss}, the
coupling of the electrons to the bath degrees of freedom is
completely specified by the bath spectral function
\begin{equation}
   J(\omega) = \pi   \sum_{n}
\lambda_{n}^{2} \delta\left( \omega -\omega_{n} \right) \ .
\end{equation}

The spectral function characterizes the bath and the system-bath
coupling, and can be related to the classical reorganization energy
\cite{Weiss} (classical in terms of boson degrees of freedom) which
measures the energy relaxation that follows a sudden electronic
transition. The one-electron transfer and the correlated
two-electron transfer are associated with reorganization energies
$E_{\alpha 1}$ and $E_{\alpha 2}$, respectively. For a
single-electron transfer, e.g., $D^{2-}A\rightarrow D^-A^{-}$ the
reorganization energy $E_{\alpha 1}$ is given by \cite{Weiss}
\begin{equation}
   E_{\alpha 1} =   \sum_n \frac{\lambda_n^2}{ \omega_n}
= \int_0^\infty \frac{d\omega}{\pi} \frac{ J(\omega)}{\omega} \ ,
\label{reo}
\end{equation}
and the corresponding energy for a correlated two-electron transfer
($D^{2-}A \rightarrow DA^{2-}$) is $ E_{\alpha 2} = 4 E_{\alpha 1}$.

The model we are considering here is completely specified by the
parameters $\Delta$, $\alpha$, $\tilde{U}$, $\varepsilon$ and the
bosonic spectral function. In the molecular electron transfer
problem the latter function reflects intramolecular vibrations and
the solvent (e.g., water or protein) or environment. Its solvent
component can be estimated from the solvent dielectric properties or
a classical molecular dynamics simulation. In the present paper we
assume an Ohmic bath model:

\begin{eqnarray}
J(\omega) = \left \{   \begin{array}{c@{\;\;:\;\;}c} 2 \pi \alpha
\omega  &  0 < \omega < \omega_c \ ,
\\ 0 & {\rm otherwise} \ . \end{array}  \right.
\end{eqnarray}
with a cut-off at energy $\omega_c$. This choice yields the
reorganization energy $E_{\alpha 1}= 2 \alpha \omega_c$ and the
energy shift $\tilde{U}_{\rm eff}=\tilde{U}-2 \alpha \omega_c$. All
parameters and physical quantities are defined in units of
$\omega_c$. Its order of magnitude for the intermolecular mode
spectrum of a polar solvent is ~0.1eV.


\section{The High-Temperature limit: Marcus theory \label{section_toymodel}}
\begin{figure}
\vspace*{0cm} \epsfxsize=6cm \centerline{\epsffile{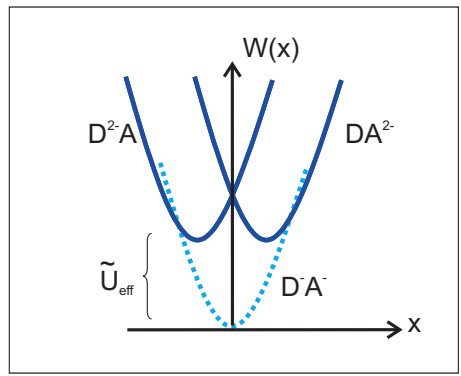}}
 \caption{Potential surfaces for the different states of the model in the Marcus theory for $\tilde{U}_{\rm eff}>0, \varepsilon=0$.
 The mimima of the
states $|\uparrow,\downarrow\rangle$ and $|\downarrow,\uparrow
\rangle$ ($D^-A^-$) are set to the origin while those
 parabolas that correspond
 to the doubly occupied states $|\uparrow\downarrow,0\rangle$ ($D^{2-}A$) and $|0,\uparrow\downarrow\rangle$
 ($DA^{2-}$) are shifted. Note that in the case displayed here the transfer $D^{2-}A \rightarrow
D^-A^-$ is in the ``inverted regime".}
 \label{Marcus}
\end{figure}

In the high-temperature, limit electron transfer is usually
described using Marcus theory \cite{Nitzan} as a rate process within
classical transition state theory. Extensions that take into account
the quantum nature of the nuclear motion in the weak electronic
coupling limit (the so called non-adiabatic limit) are also
available \cite{Nitzan}, however for simplicity we limit ourselves
in what follows to the classical Marcus description. The Marcus
electron transfer rate can be evaluated for any amount of
transferred charge: the latter just determines the renormalized
potential surface parameters that enter the rate expression.
Single-electron transition rates are given by

\begin{eqnarray}
k_{[D^{2-}A \rightarrow D^-A^-]}^{\rm single}\sim  \Delta^2 e^{-
\frac{\left(\varepsilon+\tilde{U}_{\rm eff}-E_{\alpha 1}\right)^2}{4
E_{\alpha 1}T}}, \label{Marcus1}
\end{eqnarray}
\begin{eqnarray}
k_{[D^{-}A^- \rightarrow D^{2-}A]}^{\rm single}\sim  \Delta^2 e^{-
\frac{\left(\varepsilon+\tilde{U}_{\rm eff}+E_{\alpha 1}\right)^2}{4
E_{\alpha 1}T}}, \label{Marcus2}
\end{eqnarray}
\begin{eqnarray}
k_{[DA^{2-} \rightarrow D^-A^{-}]}^{\rm single}\sim  \Delta^2 e^{-
\frac{\left(-\varepsilon+\tilde{U}_{\rm eff}-E_{\alpha
1}\right)^2}{4 E_{\alpha 1}T}},
\end{eqnarray}
\begin{eqnarray}
k_{[D^{-}A^- \rightarrow DA^{2-}]}^{\rm single}\sim  \Delta^2 e^{-
\frac{\left(-\varepsilon+\tilde{U}_{\rm eff}+E_{\alpha
1}\right)^2}{4 E_{\alpha 1}T}}.
\end{eqnarray}
In the case $\Delta \ll |U_{\rm eff}|$ second-order processes are
possible that involve only virtual occupations of the states
$D^-A^-$ leading to rates for an electron pair.
\begin{eqnarray}
k_{[D^{2-}A \rightarrow DA^{2-}]}^{\rm pair} \sim
\frac{\Delta^4}{\tilde{U}_{\rm
eff}^2}e^{\frac{(2\varepsilon-E_{\alpha 2})^2}{4  E_{\alpha 2} T}}.
\label{marcussuper}
\end{eqnarray}
\begin{eqnarray}
k_{[DA^{2-} \rightarrow D^{2-}A]}^{\rm pair} \sim
\frac{\Delta^4}{\tilde{U}_{\rm
eff}^2}e^{\frac{(2\varepsilon+E_{\alpha 2})^2}{4 E_{\alpha 2}T}}.
\label{marcussuper2}
\end{eqnarray}
The interplay between sequential and concerted two-electron transfer
(in the limit of a classical bath with a single mode or a single
reaction coordinate) can be seen from these expressions. In the
following we restrict ourselves to the symmetric case
($\varepsilon=0$). Starting with the initial state $D^{2-}A$, we
expect concerted two-electron transfer in the Marcus regime when the
rate $k_{[D^{2-}A \rightarrow DA^{2-}]}^{\rm pair}$ is larger than
the rate $k_{[D^{2-}A \rightarrow D^-A^-]}^{\rm single}$ of the
first step of the sequential process which is the case when
$|\tilde{U}_{\rm eff}|\gg T$ and $|\tilde{U}_{\rm eff}|\gg E_{\alpha
1}$ as well as $E_{\alpha 1}\leq T$.

In a parameter region where sequential transfer dominates the rates
$k_{[D^{2-}A \rightarrow D^-A^-]}^{\rm single}$ and $k_{[D^{-}A^-
\rightarrow DA^{2-}]}^{\rm single}$ as well as the corresponding
backward rates show a non-monotonic behavior and an inverted regime
dependent on the effective Coulomb interaction $\tilde{U}_{\rm eff}$
(see Fig. \ref{Marcus}).

For incoherent transfer processes (which may happen at large
temperatures and for a strong coupling to the bosonic bath), a
description of the population dynamics by kinetic equations
determined by the rates is given by
\begin{eqnarray}
\dot{d}_D (t)& &=-\left(k_{[D^{2-}A \rightarrow D^-A^-]}^{\rm
single}+k_{[D^{2-}A \rightarrow DA^{2-}]}^{\rm pair}\right) d_D(t) \nonumber \\
& &+
k_{[D^{-}A^- \rightarrow D^{2-}A]}^{\rm single} n_{DA}(t)+ k_{[DA^{2-} \rightarrow D^{2-}A]}^{\rm pair} d_A(t)  \ , \nonumber \\
\dot{n}_{DA} (t)& &=-\left(k_{[D^{-}A^- \rightarrow DA^{2-}]}^{\rm
single}+k_{[D^{-}A^- \rightarrow D^{2-}A]}^{\rm single} \right)
n_{DA}(t) \nonumber\\& &+ 2 k_{[D^{2-}A \rightarrow D^-A^-]}^{\rm
single} d_{D}(t)+ 2 k_{[DA^{2-} \rightarrow D^-A^-]}^{\rm single}
d_A(t)
\ , \nonumber \\
\dot{d}_A (t)& &=-\left(k_{[DA^{2-} \rightarrow D^-A^-]}^{\rm
single} +k_{[DA^{2-} \rightarrow D^{2-}A]}^{\rm pair}\right) d_A(t)
\nonumber \\ & &+ k_{[D^{-}A^- \rightarrow DA^{2-}]}^{\rm single}
n_{DA}(t)+
k_{[D^{2-}A \rightarrow DA^{2-}]}^{\rm pair} d_D(t),\nonumber\\
\label{kinetic}
\end{eqnarray}
where $d_D$ and $d_A$ are the probabilities to have two electrons on
the donor and acceptor, respectively. $n_{DA}$ is the combined
population of the states $|\uparrow,\downarrow\rangle$ and
$|\downarrow,\uparrow\rangle$. For the initial condition
$d_D(t=0)=1$ we obtain
$n_{|\uparrow,\downarrow\rangle}=n_{|\downarrow,\uparrow\rangle}$.
For the unbiased Hamiltonian ($\varepsilon=0$), $k_{[DA^{2-}
\rightarrow D^-A^-]}^{\rm single}=k_{[D^{2-}A \rightarrow
D^-A^-]}^{\rm single}$  and $k_{[D^-A^{-} \rightarrow DA^{2-}]}^{\rm
single}=k_{[D^{-}A^- \rightarrow D^{2-}A]}^{\rm single}$ must hold.

These kinetic equations can be solved  in the high-temperature
regime using the Marcus rates from above. In Sec.\ref{nonequi}, we
have used these equations to extract the low-temperature transition
rates by fitting the non-equilibrium dynamics of $d_D(t), d_A(t)$
and $n_{DA}(t)$ calculated in the incoherent regime with the
time-dependent NRG.

For $t \rightarrow \infty$ the equilibrium states $\langle d_A
\rangle_{\rm eq}$, $\langle d_D \rangle_{\rm eq}$ and $\langle
n_{DA} \rangle_{\rm eq}$ are reached, where $\langle d_D
\rangle_{\rm eq}=\langle d_A \rangle_{\rm eq}$. It follows that
$\frac{\langle d_D\rangle_{\rm eq}}{\langle n_{DA}\rangle_{\rm
eq}}=\frac{k_{[D^-A^{-} \rightarrow DA^{2-}]}^{\rm
single}}{k_{[DA^{2-} \rightarrow D^-A^-]}^{\rm single}}$ which is
according to the Marcus rates $\frac{\langle d_D\rangle_{\rm
eq}}{\langle n_{DA}\rangle_{\rm eq}}=e^{\tilde{U}_{\rm eff}/T}$.
Therefore, in the classical limit we arrive at
\begin{eqnarray}
\langle d_D \rangle_{\rm eq}^{\rm cl}=\frac{0.5}{
e^{\tilde{U}/T}+1}. \label{Marcus_pop}
\end{eqnarray}

With the help of the kinetic equations we can describe concerted
two-electron transfer, a purely sequential single-electron transfer
as well as a combined process which shows first a pair transfer
which is followed by a single-electron transfer. As long as the
single-electron transfer rates are small ($k_{[D^{2-}A \rightarrow
D^-A^-]}^{\rm single}<k_{[D^{2-}A \rightarrow DA^{2-}]}^{\rm pair}$)
and $\langle d_{D} \rangle_{\rm eq}=\langle d_{A} \rangle_{\rm
eq}\approx0.5$ the state $D^-A^-$ is only weakly populated and
$n_{DA}(t)$ is constant and close to zero. The dynamics is dominated
by an electron pair transfer. The combined process is expected if
$\langle d_{D} \rangle_{\rm eq}=\langle d_{A} \rangle_{\rm eq}<0.5$.
First the population $d_A$ rises quickly while $n_{DA}$ stays close
to zero. Later a slow increase of $n_{DA}$  to its equilibrium is
observed. For $k_{[D^{2-}A \rightarrow D^-A^-]}^{\rm
single}>k_{[D^{2-}A \rightarrow DA^{2-}]}^{\rm pair}$ the transfer
is purely sequential.

\section{The low-temperature limit: The Numerical Renormalization
  Group \label{sec_NRG} }

\begin{figure}

{\centering \includegraphics[width=8cm]{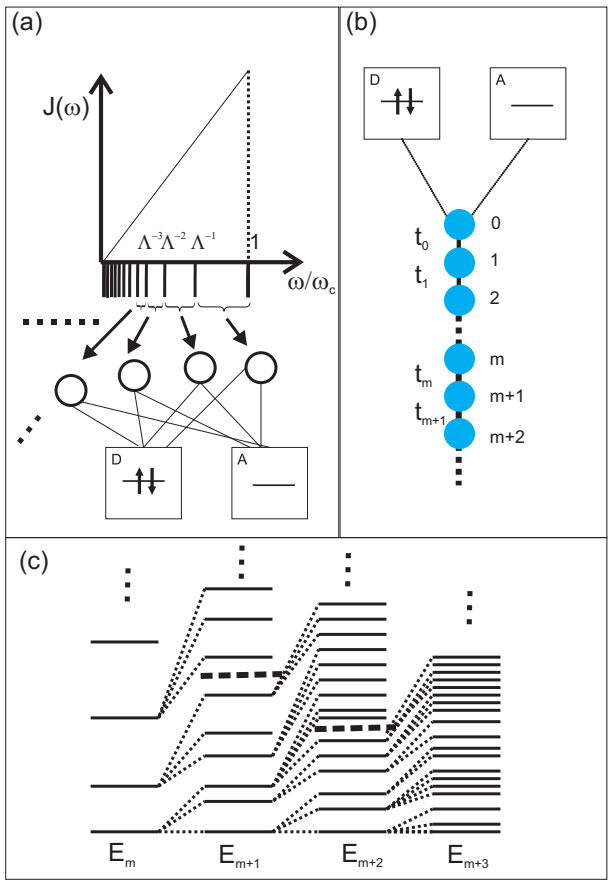}}

\caption{Scheme of the bosonic NRG. 
(a) The bosonic energy continuum is discretized on a logarithmic
mesh using a  parameter $\Lambda>1$. Only a single bosonic mode in
each  interval $[\Lambda^{-(n+1)}\omega_c,\Lambda^{-n}\omega_c]$ --
visualized by the circles -- couples directly to the electronic
subsystem. 
(b) This discretized model is mapped exactly onto a tight-binding
chain via a unitary transformation:\cite{Wil75,rev} only the first
chain site couples directly to the donor-acceptor system. The
hopping $t_{n}$ between neighboring bosonic sites decreases
exponentially with the distance from the donor-acceptor system,
i.e.~$t_n \propto \Lambda^{-n}$. 
The energy spectrum of the Hamiltonian is calculated by successively
applying the renormalization group (RG) transformation
(\ref{eqn:nrg-trafo}), diagonalizing the new Hamiltonian and
rescaling  the spectrum as depicted schematically in panel (c) for
the sequence of Hamiltonians $H_m$ to $H_{m+3}$. After each
iteration only the $N_s$ eigenstates of site $m+1$ with the lowest
energies are kept. This truncation is depicted by a horizontal
dashed line. }
\label{fig_nrg}
\end{figure}

At low temperature, the quantum generalization of the Marcus theory
replaces the classical environment by a bath of non-interacting
bosonic degrees of freedom. Very early on, the "non-adiabatic" weak
coupling limit was investigated.\cite{Levich}  The strong coupling
limit of such a model has been adressed using the non-interacting
blib approximation (NIBA) \cite{Leggett}, path integral
methods\cite{Weiss} and recently also by the numerical
renormalization group which we employ in this paper.

Originally the NRG was invented by Wilson for a fermionic bath to
solve the Kondo problem.\cite{Wil75,Kri80} The fermionic NRG is a
standard and very powerful tool to investigate complex quantum
impurity problems.\cite{rev}  The method was recently extended to
treat quantum impurities  coupled to a bosonic bath \cite{BLTV,BTV},
to a combination of fermionic and bosonic baths
\cite{GlossopIngersent2007}, and to the calculation of real-time
dynamics out of equilibrium.\cite{Costi,AS1,AS2} The
non-perturbative NRG approach has been successfully applied to
arbitrary electron-bath coupling strengths.
\cite{BLTV,BTV,GlossopIngersent2007,AndersBullaVojta2007}

\subsection{Equilibrium NRG}

The  numerical renormalization  group achieves the separation of
energy scales by logarithmic discretization of the energy continuum
into intervals $[\Lambda^{-(m+1)} \omega_c,\Lambda^{-m} \omega_c ]$,
$m\in {\cal N}_0$, defining the discretization parameter
$\Lambda>1$. Only one single mode of each interval couples directly
to the quantum impurity, indicated by the circles in
Fig.~\ref{fig_nrg}(a). This discrete representation of the continuum
is mapped onto a semi-infinite tight-binding chain using an  exact
unitary transformation. Hereby, the quantum impurity couples only to
the very first chain site as depicted in Fig.~\ref{fig_nrg}(b). The
tight-binding parameters $t_n$  linking consecutive sites of the
chain $m$ and $m+1$ fall off  exponentially as $t_m \sim
\Lambda^{-m}$. Each bosonic chain site is  viewed as representative
of an energy shell since its energy $w_m$ also decreases as $w_m
\sim \Lambda^{-m}$ establishing an energy hierarchy.  Both ensures
that mode coupling can only occur between neighboring energy shells
which is essential for the application  of the renormalization group
procedure. To this end, the renormalization  group transformation
$R[H]$ reads
\begin{eqnarray}
H_{m+1} &=& R [H_m]=\Lambda H_m+ \Lambda^{m+1}\left(t_m
a_m^{\dag} a_{m+1} \right.  \nonumber \\
 &+& \left. t_m a_{m+1}^{\dag} a_{m}    + w_m a^\dagger_{m+1}
a_{m+1} \right) \ ,\label{eqn:nrg-trafo}
\end{eqnarray}
where $H_m$ is the Hamiltonian of a finite chain up to the site $m$
-- as depicted in Fig.~\ref{fig_nrg}(b). The annihilation (creation)
operators of site $m$ are denoted by $a_m$ ($a_m^{\dag}$) and $w_m$
labels the energy of the bosonic mode of site $m$. Note that the
rescaling of the Hamiltonian $H_m$ by $\Lambda$ ensures the
invariance of the energy spectrum of fixed point Hamiltonians under
the RG transformation $R [H_m]$.

The RG transformation (\ref{eqn:nrg-trafo}) is used to set up and
iteratively diagonalize the sequence of Hamiltonians $H_n$. In the
first step, only the electronic donor-acceptor system coupling to
the single bosonic site $m=0$ is considered. It turns out to be
sufficient\cite{BLTV,BTV,rev} to include only the $N_b$ lowest lying
bosonic states, where $N_b$ takes typical values of $8-12$. The
reason for that is quite subtle: the coupling between different
sites decays exponentially and is restricted to nearest-neighbor
coupling by construction, both essential for the RG procedure. In
each successive step (i) a finite number of  $N_b$ bosonic states of
the next  site $m+1$ are added, (ii) the Hamiltonian matrices are
diagonalized and (iii) only the lowest $N_s$ states are retained in
each iteration.  The discarding of high-energy states is justified
by the Boltzmannian form of the equilibrium density operator when
simultaneously the temperature is lowered  in each iteration step to
the order $T_m\propto \Lambda^{-m}w_c$.

To illustrate the procedure, the lowest-lying energies of the
Hamiltonian $H_m$ to $H_{m+3}$ are schematically depicted in panel
(c) of Fig.~\ref{fig_nrg}. We typically use  $N_b \geq 8$ and keep
about $N_s=100$ states after each iteration using a discretization
parameter $\Lambda=2$.

Denoting the set of low-lying eigenstates by $|r\rangle_N$ and the
corresponding eigenvalues $E_r(N) \propto O(1)$ at iteration $N$,
the equilibrium  density matrix $\rho_0$ is given\cite{rev} by
\begin{eqnarray}
  \rho_0 &=& \frac{1}{Z_N} \sum_{r} e^{-\bar \beta E^N_r} |r\rangle_N
  {}_N\langle r| \ ,
\label{eqn:rho0}
\end{eqnarray}
where $Z_N = \sum_r e^{-\bar \beta E^N_r}$ and $\bar\beta$ are of
the order $O(1)$, such that $T_N = w_c \Lambda^{-N}/\bar \beta$. The
thermodynamic expectation value of each local observable $\hat O$ is
accessible at each temperature $T_N$ by the trace
\begin{eqnarray}
  \langle \hat O \rangle_{\rm eq} &=& \mbox{Tr}\left[\rho_0 \hat
    O\right]
= \frac{1}{Z_N} \sum_{r} e^{-\bar \beta E^N_r} {}_N\langle r| \hat O
|r\rangle_N
 \label{ddT}
\, .
\end{eqnarray}
The procedure described above turns out to be very accurate because
the couplings $t_m$ between the bosonic sites along the chain are
falling off exponentially, so that  the rest of the semi-infinite
chain contributes only perturbatively\cite{Wil75,rev} at each
iteration $m$ while contributions from the discarded high-energy
states are exponentially suppressed by the Boltzmann factor.

\subsection{Time dependent NRG}

While the equilibrium properties are fully determined by the energy
spectrum of the Hamiltonian, the non-equilibrium dynamics requires
two conditions: the initial condition encoded in the many-body
density operator $\rho_0$ and the Hamiltonian $H^{\rm f}$ which
governs its time-evolution. For a time-independent Hamiltonian, the
density operator evolves according to $\hat{\rho} (t>0)=e^{-iH^{\rm
f} t} \rho_0 e^{iH^{\rm f} t}$. All time-dependent expectation
values $ \langle \hat O \rangle(t)$ are given by
\begin{eqnarray}
  \langle \hat O \rangle(t) &=& \mbox{Tr}\left[\hat{\rho(t)} \hat
    O\right]
= \mbox{Tr}\left[e^{-iH^{\rm f} t} \rho_0 e^{iH^{\rm f} t} \hat
O\right] \; . \label{eqn:o-real-time}
\end{eqnarray}

We obtain the density operator $\rho_0$ from an independent NRG run
using a suitable initial Hamiltonian $H^{\rm i}$. By choosing
appropriate parameters in $H^{\rm i}$, we prepare the system such
that (for the calculations presented in this paper) the two
electrons are located on the donor site and  the acceptor site is
empty.

In general, the initial density operator $\rho_0$ contains states
which are most likely superpositions of excited states of $H^{\rm
f}$. For the calculation of the real-time dynamics of
electron-transfer reactions it is therefore not sufficient to take
into account  only the retained states of the Hamiltonian $H^{\rm
f}$ obtained from an NRG procedure. The recently developed
time-dependent NRG (TD-NRG)\cite{AS1,AS2} circumvents this problem
by including contributions from all states. It turns out that the
set of all discarded states eliminated during the  NRG procedure
form a complete basis set\cite{AS1,AS2} of the Wilson chain which is
also an approximate eigenbasis of the Hamiltonian. Using this
complete basis, it was shown\cite{AS1,AS2} that
eq.~(\ref{eqn:o-real-time}) transforms into the central equation of
the TD-NRG for the temperature $T_N$
\begin{eqnarray}
\langle \hat{O} \rangle (t) &=&
        \sum_{m = 0}^{N}\sum_{r,s}^{\rm trun} \;
        e^{i(E_{r}^m -E_{s}^m)t}
        O_{r,s}^m \rho^{\rm red}_{s,r}(m) \; ,
\label{eqn:time-evolution}
\end{eqnarray}
where $O_{r,s}^m = \langle r;m| \hat{O}| s;m \rangle$ are the matrix
elements of any operator $ \hat{O}$ of the electronic subsystem at
iteration $m$, and $E_{r}^m,E_{s}^m$ are the eigenenergies of the
eigenstates $|r;  m\rangle$  and $|s;  m\rangle$ of $H^{\rm f}_m$.
At each iteration $m$, the chain is formally partitioned into a
``system'' part on which the Hamiltonian $H_m$ acts exclusively and
an environment part formed by the bosonic sites $m+1$ to $N$.
Tracing out these environmental degrees of freedom $e$ yields  the
reduced density matrix\cite{AS1,AS2}
\begin{equation}
\rho^{\rm red}_{s,r}(m) = \sum_{e}
          \langle s,e;m|\rho_{0} |r,e;m \rangle
\label{eqn:reduced-dm-def}
\end{equation}
at iteration $m$, where $\rho_0$ is given by (\ref{eqn:rho0}) using
$H^{\rm i}$. The restricted sum $\sum^{\rm trun}_{r,s}$ in
eq.~(\ref{eqn:time-evolution}) implies that at least one of the
states $r$ and $s$ is discarded at iteration $m$. Excitations
involving only kept states contribute at a later iteration and must
be excluded from the sum.

As a consequence,  {\em all} energy shells $m$ contribute to the
time evolution: the short time dynamics is governed by the high
energy states while the long time behavior is determined by the low
lying excitations. Dephasing and dissipation is encoded in the phase
factors $ e^{i(E_{r}^m -E_{s}^m)t}$ as well as the reduced density
matrix $\rho^{\rm red}_{s,r}(m)$.

Discretizing the bath continuum will lead to finite-size
oscillations of the real-time dynamics around the continuum solution
and deviations of expectation values from the true equilibrium at
long time scales. In order to separate the unphysical finite-size
oscillations from the true continuum behavior, we average over
different bath discretization schemes using Oliveira's $z$-averaging
(for details see Refs.~\onlinecite{oliveira,AS2}). We average over 8
different bath discretizations in our calculation.

\section{Equilibrium properties \label{equil}}

In order to gain a better understanding of the non-equilibrium
dynamics presented in Sec.~\ref{nonequi}, we briefly summarize the
equilibrium properties of the model given by Eq.~(\ref{eq:ebm}). It
has been analyzed in Ref.~[\onlinecite{TTB1}], where self-trapping
(localization) in the single and two-electron subspace was found.

\begin{figure}
\vspace*{0cm} \hspace*{0cm} \epsfxsize=9cm
\centerline{\epsffile{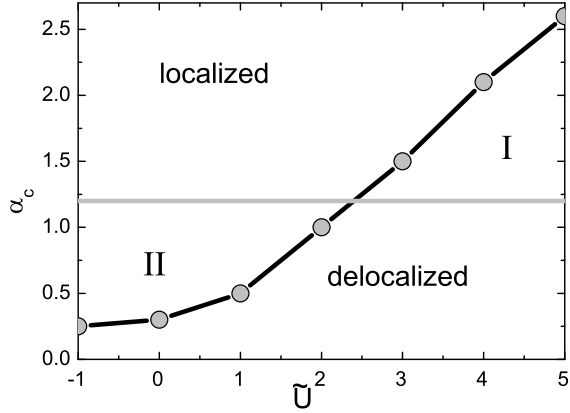}}
 \caption{Zero-temperature phase diagram of the model
eq.~(\ref{eq:ebm}) for $\varepsilon=0$ and $\Delta=0.1 \omega_c$.
The critical dissipation strength $\alpha_c$ is plotted as a
function of $\tilde{U}$ in the two-particle subspace (black line)
and in the single-particle subspace (grey line), respectively.}
\label{phase}
\end{figure}

We start with the phase diagram of the two-site model, as shown in
Fig. \ref{phase}. Only for $\varepsilon=0$ a quantum phase
transition of Kosterlitz-Thouless type separates a localized phase
for $\alpha>\alpha_c$ from a delocalized phase for
$\alpha<\alpha_c$. We plot the phase boundaries between localized
and delocalized phases in the $\alpha$-$\tilde{U}$-plane, both for
single- and two-electron subspaces (grey and black line in
Fig.~\ref{phase}, respectively).

For the single-electron subspace, the Coulomb repulsion is
irrelevant, and the phase boundary does not depend on $\tilde{U}$.
The value of the critical coupling  strength, $\alpha_{\rm c}$, is
identical to those of the corresponding spin-boson model. The
critical value\cite{Leggett,BLTV} of $\alpha_c$  depends on the
tunneling rate $\Delta$  and reaches $\alpha_c=1$  for $\Delta\to
0$.

The phase boundary for the two-electron subspace does depend on
$\tilde{U}$, which has drastic consequences for the electron
transfer process. Imagine that, by a suitable choice of parameters,
the system is placed between the two phase boundaries above the
single-electron (grey line) and below the two-electron phase
boundary (black line) in the area indicated by I in Fig.
\ref{phase}. Then the system would be in the localized phase in the
single-electron subspace. However, one additional second electron
immediately places the system in the delocalized phase, and one or
even both electrons can be transfered. Similarly, a second electron
added to the system in the parameter regime of area II shows the
opposite behavior: both electrons get localized although a single
electron could be transfered.

Note the different values of the $\alpha_{\rm c}$'s even for
$\tilde{U}=0$ in the single and the two-electron subspace: the
coupling of the donor/acceptor  system to the bath induces an
effective attractive Coulomb interaction $\tilde{U}_{\rm
eff}=-2\alpha \omega_c$ between the electrons. On the localized side
of the transition, the electron tunneling $\Delta$ is renormalized
to zero, so that an electron transfer is clearly absent in this
regime. This statement holds only for Ohmic dissipation, on which we
focus here; deep in the sub-ohmic regime, coherent oscillations have
been recently observed even in the localized phase, see
Ref.~\onlinecite{AndersBullaVojta2007}.

\begin{figure}
\vspace*{0cm} \epsfxsize=10cm \centerline{\epsffile{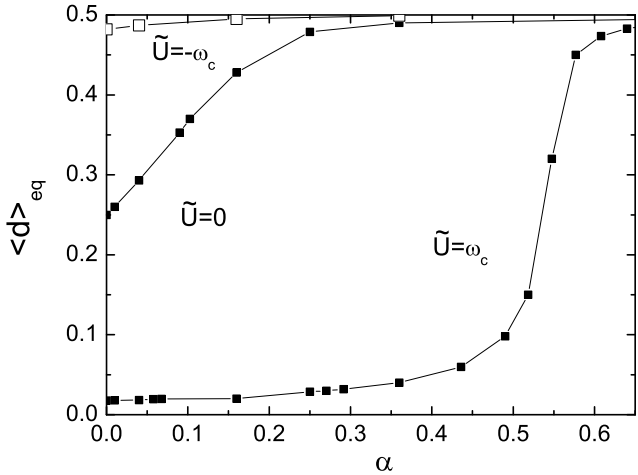}}
 \caption{Low-temperature equilibrium probability for double occupancy of
 donor and acceptor
$\langle d\rangle_{\rm eq}$ for $\Delta=0.1 \omega_c $,
$\varepsilon=0$, as a function of $\alpha$ for
$\tilde{U}=-\omega_c,0$ and $\omega_c$. In the limit of $\alpha=0$
the dependence of $\langle d \rangle_{\rm eq}$ on $\Delta$ and
$\tilde{U}$ is given analytically in eq.~(\ref{double_eq}).}
 \label{dequ}
\end{figure}

Figure \ref{dequ} shows results for the double occupation
probability as a function of the electron-bath coupling $\alpha$ for
different $\tilde{U}$ calculated with the equilibrium NRG. For the
symmetric model considered here, the equilibrium probabilities for
the double occupation on donor and acceptor sites are equal:
$\langle \hat d_{\rm A}\rangle_{\rm eq} = \langle \hat d_{\rm
D}\rangle_{\rm eq} \equiv \langle d\rangle_{\rm eq}$ using the
observables defined in Eq.~(\ref{eqn:observables}).  The probability
of having two electrons at different sites is given by $\langle \hat
n_{\rm
  DA}\rangle_{\rm eq}=1- 2\langle d \rangle_{\rm eq}$.

The average double  occupancy  $\langle  d \rangle_{\rm eq}$
decreases with increasing effective Coulomb repulsion  $\tilde{U}$
and increases with increasing $\alpha$. This can  understood in
terms of the effective Coulomb interaction $\tilde{U}_{\rm eff}=
\tilde{U}-2  \alpha \omega_c$, renormalized due to the coupling to
the  bosonic bath.

The delocalization/localization phase transition  occurs when
$\langle d \rangle_{\rm eq}\rightarrow 0.5$, as can be seen by
comparing Fig.~\ref{phase} and Fig.~\ref{dequ}. For  $\tilde{U}_{\rm
  eff}< 0$ and $\tilde{U}_{\rm
  eff}\gg \Delta$,  we are able to project out the $D^{-}A^{-}$ excited
states. Then our model maps on a spin-boson model with an effective
hopping $\Delta/\tilde{U}_{\rm
  eff}^2$ between the states $D^{2-}A$ and $DA^{2-}$. The
dynamics will be governed by electron pairs if $D^{2-}A$ or
$DA^{2-}$ are the initial states.

The double occupancy $\langle d\rangle_{\rm eq}$ is calculated
analytically for $\alpha=0$ and arbitrary $\Delta$ and $\tilde U$.
For $T\to 0$, the $\langle d\rangle_{\rm eq}$ approaches
\begin{eqnarray}
\langle d\rangle_{\rm eq}=\frac{4 \Delta^{2}}{\sqrt
{\tilde{U}^{2}+16 {\Delta}^{2}}\, (\tilde{U}+\sqrt {\tilde{U}^{2}+16
{\Delta}^{2}})} \label{double_eq}\ .
\end{eqnarray}
while in the opposite limit, $T\to\infty$, we obtain $\langle d
\rangle_{\rm eq} \to 0.25$. The low-temperature limit
(\ref{double_eq}) is included as end-points of the curves in
Fig.~\ref{dequ}.

\begin{figure}
\vspace*{0cm} \hspace*{0cm} \epsfxsize=10cm
\centerline{\epsffile{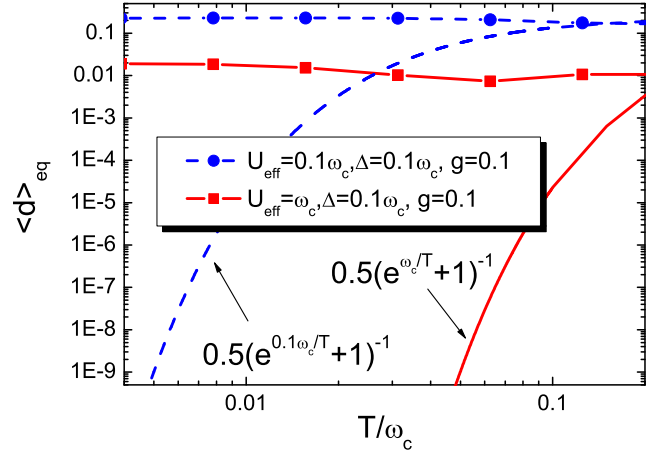}}
 \caption{Equilibrium probability for double occupancy of donor and acceptor
$\langle d\rangle_{\rm eq}$ as a function of temperature $T$ for
$\Delta=0.1 \omega_c$, $\tilde{U}_{\rm eff}=0.1\omega_c$,
$\alpha=0.04$
 (circles, dashed line) and or $\Delta=0.1 \omega_c$, $\tilde{U}_{\rm
eff}=\omega_c$, $\alpha=0.04$ (squares, solid line). For comparison
the "high temperature" result eq. (\ref{Marcus_pop}) is shown for
$\tilde{U}_{\rm eff}=0.1\omega_c$ (dashed line) and $\tilde{U}_{\rm
eff}=\omega_c$ (solid line).} \label{du_temp}
\end{figure}

Let us now turn to the temperature dependence of $\langle d
\rangle_{\rm eq}$. Figure \ref{du_temp} shows results for
temperatures between $T =0.004 \omega_c$ and $T = 0.2 \omega_c$ for
several choices of model parameters. Our calculations imply an
independent check of the correct $t \rightarrow \infty$ behavior in
the next section. Additionally, we can make connection  to the
high-temperature results of section \ref{section_toymodel}. For
temperatures $T\ll \tilde{U}_{\rm eff}$ the double occupancy
$\langle d\rangle_{\rm eq}$ is constant as expected from quantum
statistics but deviates drastically from the predictions of the
Marcus theory given by Eq.~(\ref{Marcus_pop}). The double occupancy
$\langle d_D\rangle_{\rm eq}$ calculated with the NRG approaches the
value $0.5/(1+e^{\tilde{U}_{\rm eff}/T})$ for $\tilde{U}_{\rm eff}
\approx T$. This result indicates that for $\tilde{U}_{\rm eff}>T$
Marcus theory is not applicable while low temperature methods like
the NRG are valid.

\section{Non-Equilibrium Dynamics \label{nonequi}}

We employ the time-dependent NRG to evaluate the low-temperature
time evolution of the local occupancies using
Eq.~(\ref{eqn:time-evolution}) and investigate the influence of
different Coulomb interactions $\tilde{U}$, single-electron hopping
matrix elements $\Delta$, couplings between the electronic system to
the bosonic bath $\alpha$ and temperatures $T$ between $T=3 \cdot
10^{-8}\omega_c$ and  $T=0.125 \omega_c$. The donor/acceptor
sub-system is initially prepared in a state with the two electrons
placed on the donor site and evolves according to Hamiltonian
(\ref{eq:ebm}). We calculate the time-dependent expectation values
$d_{\rm D}(t)= \langle \hat d_{D} \rangle(t)$, $d_{\rm A}(t)=
\langle \hat d_{A} \rangle(t)$ and $n_{\rm DA}(t) = \langle \hat
n_{DA} \rangle(t)$ using Eq.~(\ref{eqn:time-evolution}). These
expectation values are related at any time by the completeness
relation $d_D(t)+d_{A}(t)+n_{DA}(t)=1$. The time evolution of
$n_{DA}(t)$ serves as criterion to  distinguish between direct
two-electron transfer and two consecutive one-electron steps.  If
$n_{DA}(t)$ remains close to zero or stays constant throughout the
electron transfer process, the two states $D^{-}A^-$ are only
virtually occupied, and concerted two-electron transfer is observed.
A significant increase of $n_{DA}(t)$ as function of time is taken
as an indication of step-by-step single-electron transfer.

In the absence of the electron-boson coupling ($\alpha=0$), the
dynamics is fully determined by the dynamics of the four eigenstates
of $H_{el}$. In the limit  $|\tilde{U}| \gg \Delta$ we obtain
\begin{eqnarray}
& & d_{D;A}(t) \approx \frac{1}{2}-\frac{2 \Delta^2 }{ \tilde{U}^2}
+ \frac{2 \Delta^2 }{ \tilde{U}^2} \cos\left(\tilde{U} t \right) \pm
\frac{1}{2} \cos\left(\frac{4\Delta^2}{\tilde{U}}
 t \right) \ ,\nonumber \\
& & n_{DA}(t) \approx \frac{4 \Delta^2 }{ \tilde{U}^2} - \frac{4
\Delta^2 }{ \tilde{U}^2} \cos\left(\tilde{U} t \right) ,
\label{occ_tU}
\end{eqnarray}
while in the limit of $\tilde{U}=0$
\begin{eqnarray}
& &d_{D;A}(t)=\frac{3}{8}+\frac{1}{8} \cos(4 \Delta t)\pm
\frac{4}{8}
\cos(2 \Delta t) \ , \nonumber \\
& & n_{DA}(t)=\frac{2}{8}-\frac{2}{8} \cos(4 \Delta t) . \label{occ}
\end{eqnarray}

\begin{figure}
\vspace*{0cm} \epsfxsize=7cm \centerline{\epsffile{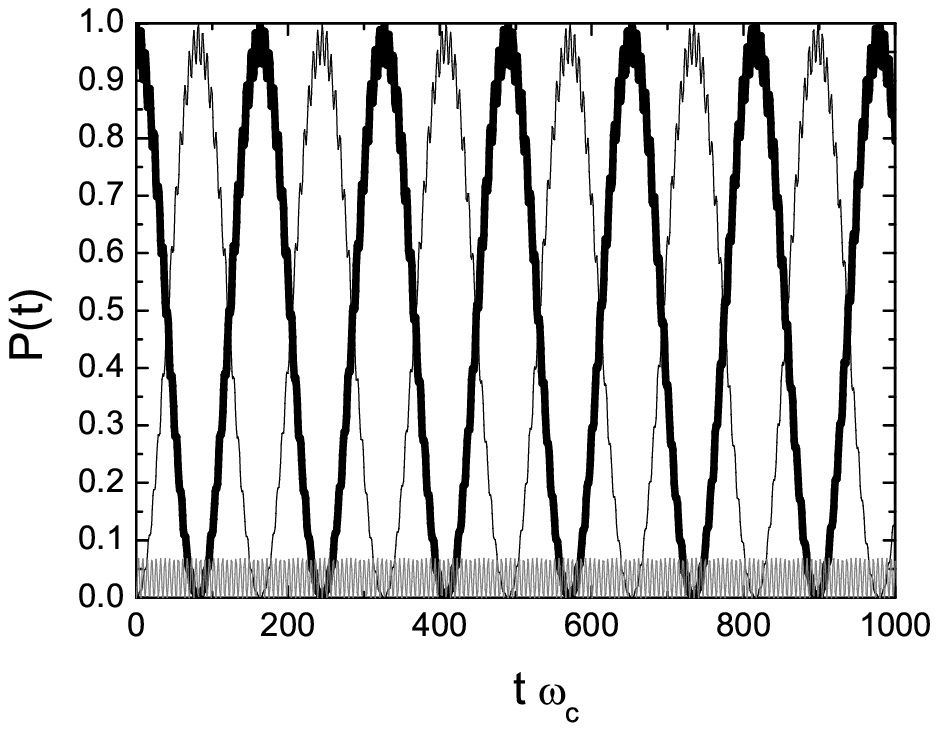}}
\centerline{\epsffile{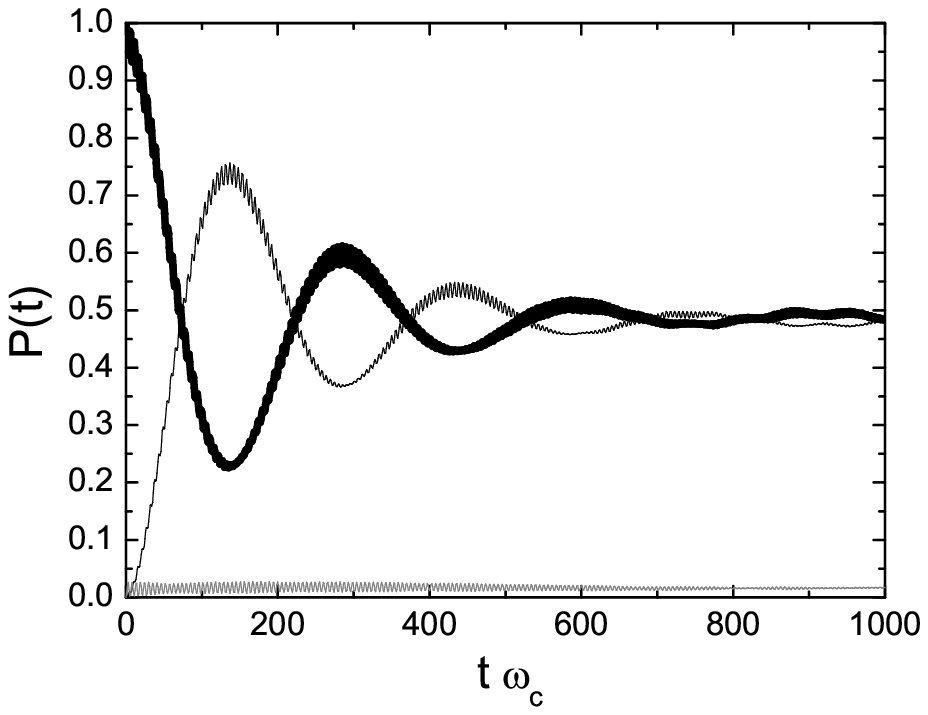}} \centerline{\epsffile{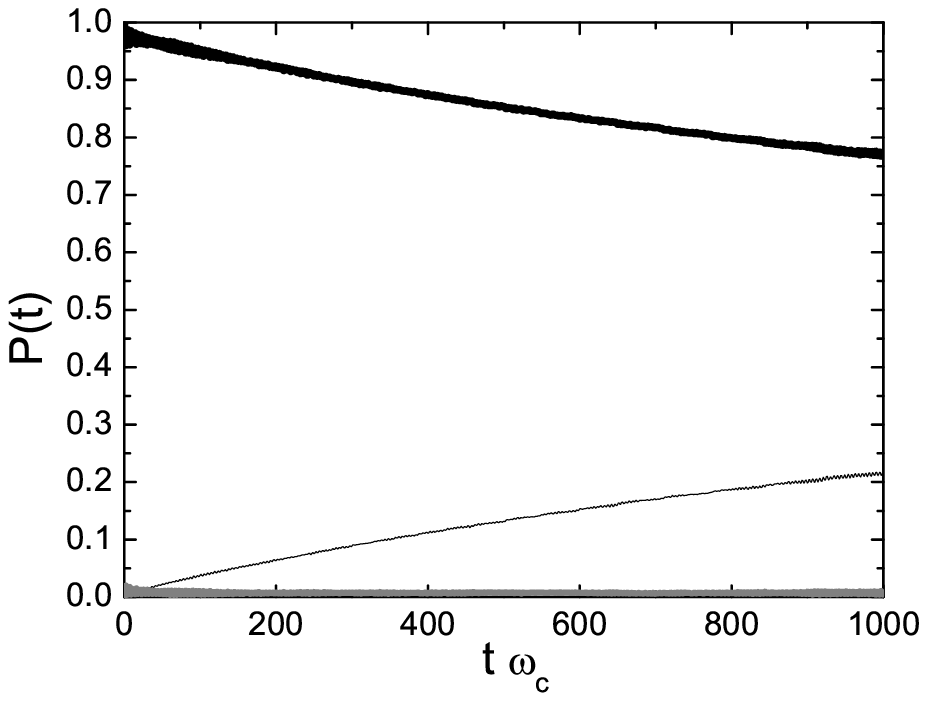}}
 \caption{Low-temperature population probabilities $P(t)=$ $d_D (t)$ (thick black line), $d_A(t)$
 (thin black line) and $n_{DA}(t)$ (gray line) as functions of time. The
parameters are $ \tilde{U}=-\omega_c$, $\Delta=0.1 \omega_c$,
$\varepsilon=0$ and $T= 3 \cdot 10^{-8} \omega_c$. The coupling
$\alpha$ increases from the upper panel $\alpha=0$ $(\tilde{U}_{\rm
eff}=-\omega_c)$, the middle panel $\alpha=0.04$ ($\tilde{U}_{\rm
eff}=-1.08\omega_c$) to the lower panel $\alpha=0.16$
($\tilde{U}_{\rm eff}=-1.32\omega_c$). }
 \label{finiteV}
\end{figure}

A finite value of the coupling, $\alpha \neq 0$, gives rise to
damping of those coherent oscillations.  Furthermore, the Coulomb
interaction is renormalized to $\tilde{U}_{\rm eff}=\tilde{U}-2
\alpha \omega_c$. For $\tilde{U}_{\rm eff}<0$, the states $D^{2-}A$
and $DA^{2-}$ are energetically favored. The two intermediate states
$D^-A^-$ are only virtually occupied for $|\tilde{U}_{\rm eff}| \gg
\Delta,T$, similar to the superexchange process.\cite{Jortner} This
regime can be described by a spin-boson model with an effective
interstate coupling $\Delta_{{\rm eff}}\approx
4\Delta^2/\tilde{U}_{\rm eff}$. The spin-boson model has three
dynamical regimes.\cite{Leggett}.

For $\alpha$ smaller than some characteristic value it exhibits
damped coherent oscillations between the two states. If $\alpha$ is
larger than this value the oscillations disappear and the kinetics
is dominated by a relaxation process. Here, rates can be defined and
the population probabilities can be fitted with the kinetic
equations (\ref{kinetic}). For a further increase of $\alpha$, the
electronic system shows the onset of localization (for $T
\rightarrow 0$) and does not evolve towards the other (acceptor)
site.

In Fig.~\ref{finiteV} we plot the low-temperature population
probabilities $d_D(t)$, $d_A(t)$ and $n_{DA}(t)$ for
$\tilde{U}=-\omega_c$, $\Delta=0.1 \omega_c$, $T=3 \cdot 10^{-8}
\omega_c$ and different couplings $\alpha$. For $\alpha=0$ (upper
panel) the oscillations have two frequencies (see
eq.~(\ref{occ_tU})). The electron pair oscillates from donor to
acceptor with the small frequency $4\Delta^2/\tilde{U} $ whereas the
fast oscillations with frequency $\tilde{U}$ characterize the
virtual occupation of the high lying states ($D^-A^-$). An increase
of $\alpha$ leads to damping of the oscillations (middle panel) and
relaxation (lower panel). At about $\alpha=0.3$ the electron pair
gets self-trapped and the system shows a phase transition to the
localized phase at T=0 (see Fig.~\ref{phase}). The configuration
$D^-A^-$ is seen not to be involved in the dynamics as $n_{DA}$ is
very small and without ascending slope. Since $\Delta \ll
\tilde{U}_{\rm eff}$ the state $D^-A^-$ cannot be populated as long
$\tilde{U}_{\rm eff} \gg T$.

\begin{figure}
 \vspace{0cm}\hspace*{0cm} \epsfxsize=10cm
\centerline{\epsffile{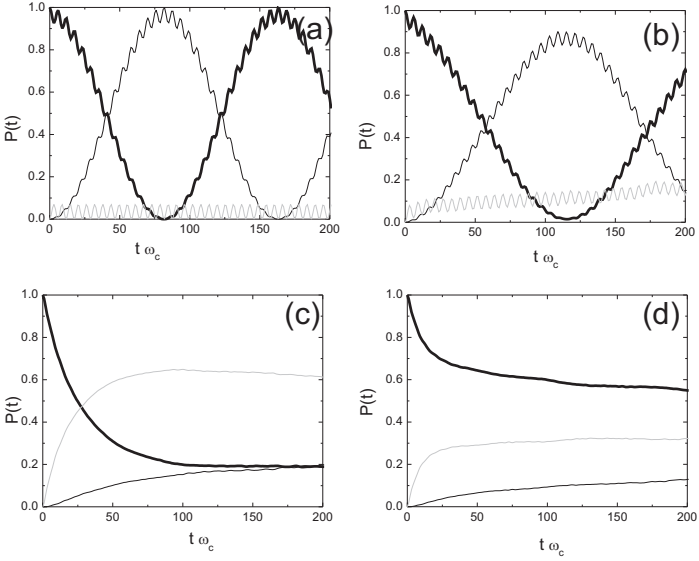}}
 \caption{Low-temperature population probabilities $P(t)=$ $d_D (t)$ (thick black line), $d_A(t)$
 (thin black line) and $n_{DA}(t)$ (gray line) as functions of time.
The parameters are $\tilde{U}=\omega_c$, $\Delta=0.1 \omega_c$,
$\varepsilon=0$ and $T=3 \cdot 10^{-8}\omega_c$. The coupling to the
bosonic bath increases from panel (a) $\alpha=0 (\tilde{U}_{\rm
eff}=\omega_c)$, panel (b) $\alpha=0.02$ $(\tilde{U}_{\rm eff}=0.96
\omega_c)$, panel (c) $\alpha=0.52$ $(\tilde{U}_{\rm eff}=-0.04
\omega_c)$ to panel (d) $\alpha=0.55$ $(\tilde{U}_{\rm eff}=-0.1
\omega_c)$. } \label{figU1t01a0}
\end{figure}

\begin{figure}
 \vspace{0cm}\hspace*{0cm} \epsfxsize=7cm
\centerline{\epsffile{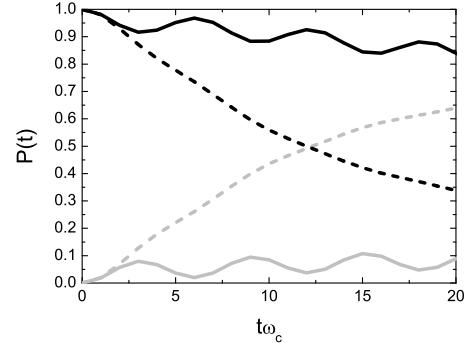}}
 \caption{Low temperature population probabilities $P(t)=$ $d_D (t)$ (black line) and $n_{DA}(t)$ (gray line) as functions of time
 for $\alpha=0.04$ (full line), $\alpha=0.36$ (dashed line).
The effective energy difference between the states $D^{2-}A$ and
$D^-A^-$ is kept constant $\tilde{U}_{\rm eff}=\omega_c$. The other
parameters are $\Delta=0.1 \omega_c$ and $\varepsilon=0$.
\label{figUeff1}}
\end{figure}

\begin{figure}
\hspace*{0cm} \vspace{0cm} \epsfxsize=10cm
\centerline{\epsffile{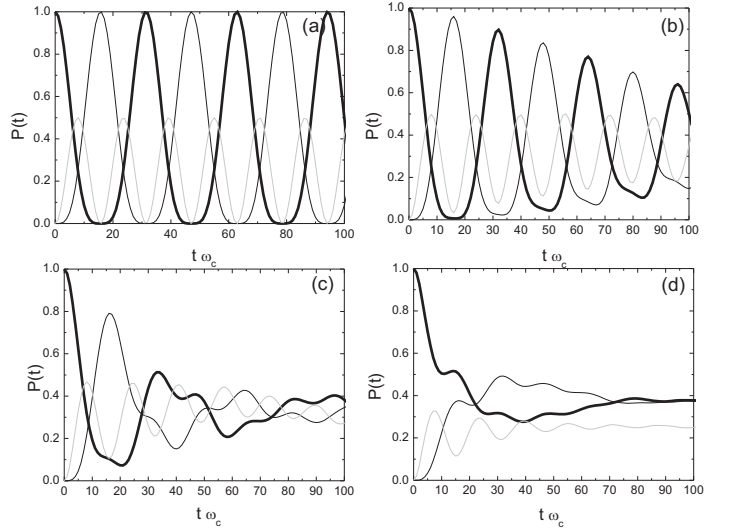}}
 \caption{Low temperature population probabilities $P(t)=$ $d_D (t)$ (thick black line), $d_A(t)$
 (thin black line) and $n_{DA}(t)$ (gray line) as functions of time. The
parameters are $\tilde{U}=0, \Delta=0.1 \omega_c$, $\varepsilon=0$
and $T = 3 \cdot 10^{-8}\omega_c$. (a) $\alpha=0 (\tilde{U}_{\rm
eff}=0)$, (b) $\alpha=0.01 (\tilde{U}_{\rm eff}=-0.02\omega_c)$, (c)
$\alpha=0.04 (\tilde{U}_{\rm eff}=-0.08 \omega_c)$ and (d)
$\alpha=0.1 (\tilde{U}_{\rm eff}=-0.2 \omega_c)$. \label{U0t01b}}
\end{figure}

A more complicated behavior is expected within the four accessible
electronic states when $\tilde{U}_{\rm eff} \gg \Delta > 0$. In this
case the delocalized states $D^-A^-$ have the lowest energy, and
sequential transfer is required to reach the equilibrium state. Pair
transfer occurs  on a smaller time scale. Thus, a combined pair and
sequential transfer on two different time scales governs the
dynamics for these parameters.

The four panels in Fig.~\ref{figU1t01a0} depict the time evolution
of the occupation probabilities $d_D(t)$, $d_A(t)$ and $n_{DA}(t)$
for $\tilde{U}=\omega_c$, $\Delta=0.1\omega_c$ and four different
values of $\alpha$: $\alpha=0,0.02,0.52,0.54$. The undamped coherent
oscillations of panel (a) decay exponentially for small damping
depicted in panel (b). Increasing $\alpha$ further yields a finite
population of the states $D^-A^-$: sequential transfer becomes the
main process, as shown in panels (c) and (d). 
The crossover from a combined  pair-transfer and slow
single-electron transfer (panel(b)) to purely sequential transfer
(panel (c) and panel (d)) with a complex dynamics is due to a
combined effect of dissipation and decrease of the effective energy
difference between the donor/acceptor states. An even larger
$\alpha$ leads to a negative $\tilde{U}_{\rm
  eff}$ and a very slow transfer until the onset of localization
at $\alpha_c$ which is not shown here.

\begin{figure}
\vspace*{0cm} \hspace*{0cm} \epsfxsize=6cm
\centerline{\epsffile{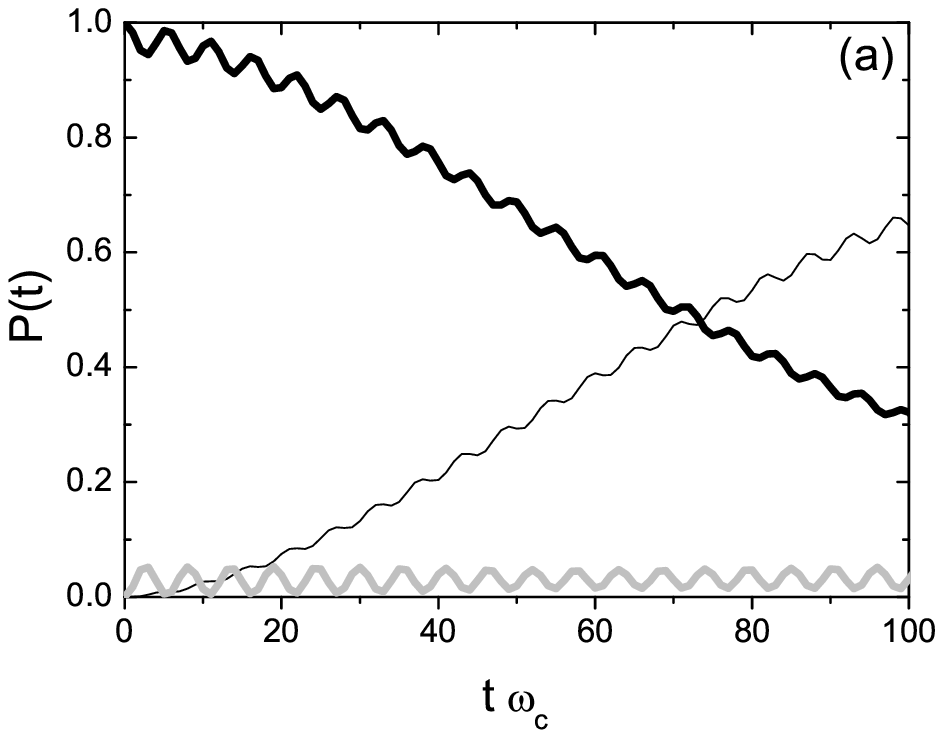}} \centerline{\epsffile{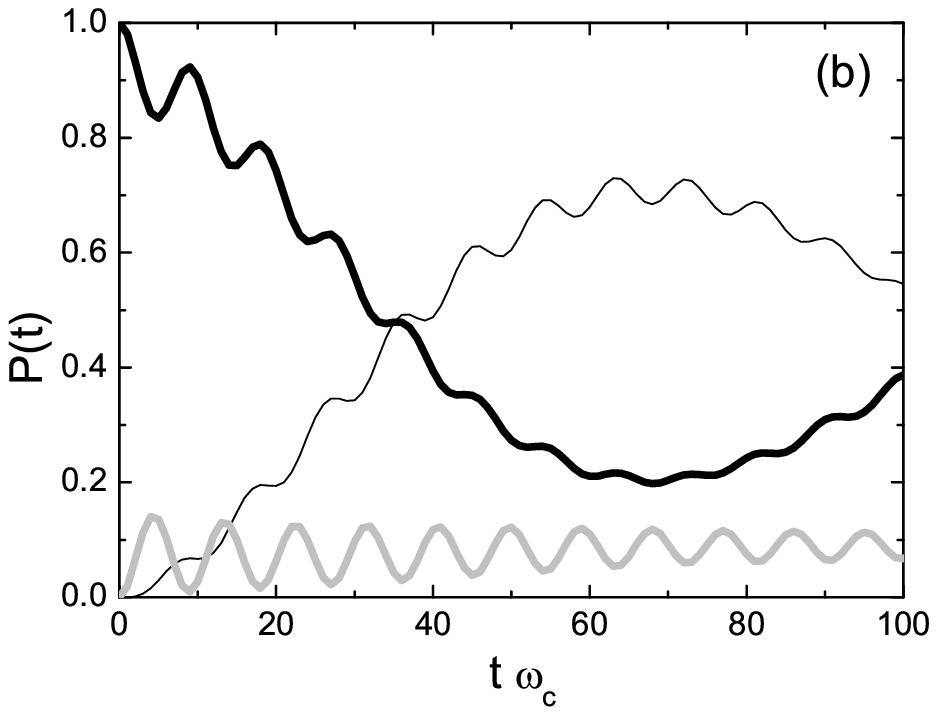}}
\centerline{\epsffile{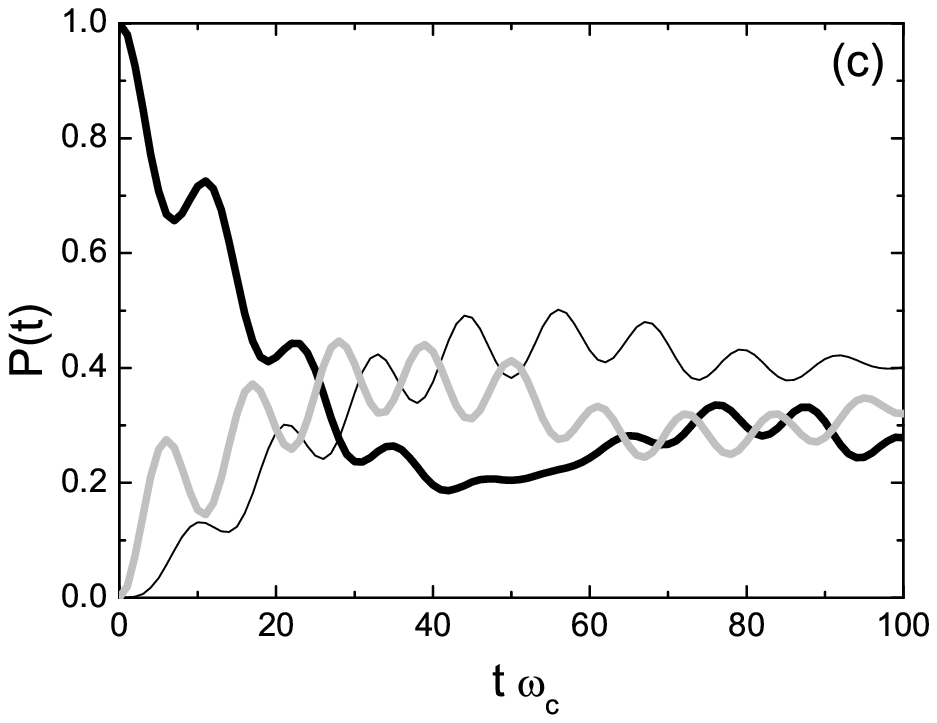}} \centerline{\epsffile{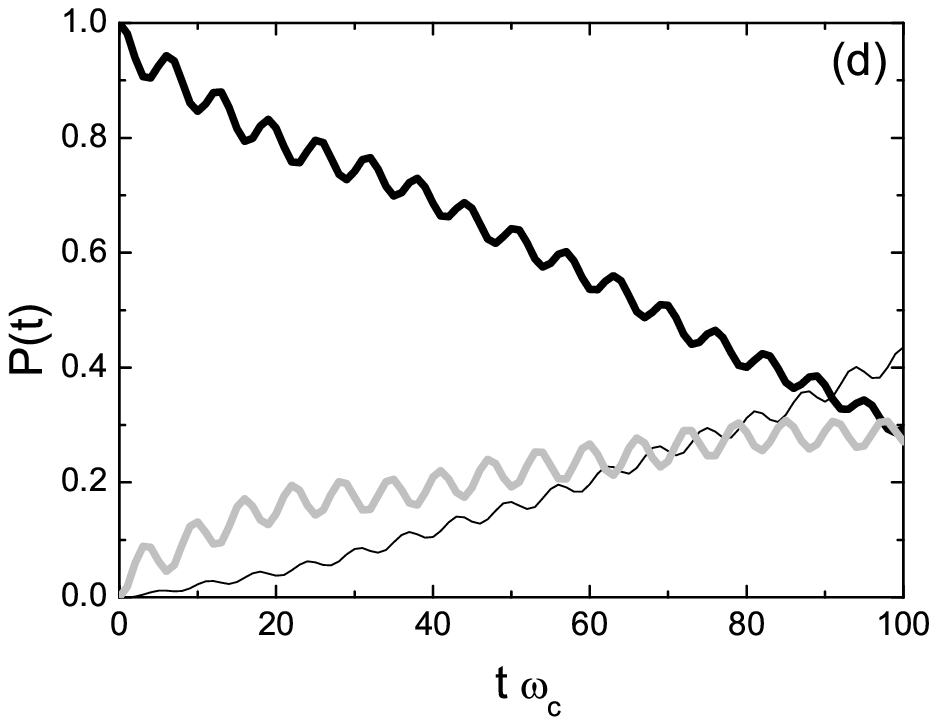}}
 \caption{ Low temperature population probabilities $P(t)=$ $d_D (t)$ (thick black line), $d_A(t)$
 (thin black line) and $n_{DA}(t)$ (gray line) as functions of time. The parameters are $\alpha=0.04$ and $\tilde{U}=-\omega_c$ (panel (a)),
$\tilde{U}=-0.5 \omega_c$ (panel (b)), $\tilde{U}=0.5 \omega_c$
(panel (c)) and $\tilde{U}=\omega_c$ (panel (d))  \label{fig_a01}}
\end{figure}
To separate the influence of dissipation from the renormalization of
$\tilde{U}$ due to the coupling to the bosonic bath we plot $d_D(t)$
and $n_{DA}(t)$ for a constant effective $\tilde{U}_{\rm
eff}=\omega_c$ and different coupling $\alpha$ in Fig.\
\ref{figUeff1}. The dynamics  changes from pair transfer with a slow
increase of the single occupancy at $\alpha=0.04$ (due to the
low-lying  states $D^-A^-$) to incoherent relaxation and sequential
transfer for $\alpha=0.36$. 
As long as $E_{\alpha 1}=2\alpha \omega_c \ll\tilde{U}_{\rm eff}$,
pair transfer is observed on a short  time-scale. For $E_{\alpha 1}
\geq \tilde{U}_{\rm eff}$ only one electron is transferred and the
system relaxes rapidly into its equilibrium state $D^-A^-$ without
any short time concerted pair transfer.

In Fig.  \ref{U0t01b} the evolution of the dynamics is shown for
$\tilde{U}=0$ and increasing $\alpha$. The doubly occupied states
are the ground states of the donor/acceptor system for finite
$\alpha$ since $\tilde{U}_{\rm   eff}=-2\alpha \omega_c<0$. With
increasing $\alpha$, the amplitude of coherent oscillations acquire
a small damping. In addition, pair transfer is favored and
$n_{DA}(t)$ decreases. The simple damped oscillations are replaced
by a much more complex dynamics comprising of strongly renormalized
oscillation frequency and a strong damping for $\alpha=0.04$. At
about $\alpha=0.36$ -- not shown here -- the critical coupling
$\alpha_c$ is reached and the system is localized.

\begin{figure}
\vspace*{0cm} \hspace*{0cm} \epsfxsize=7cm
\centerline{\epsffile{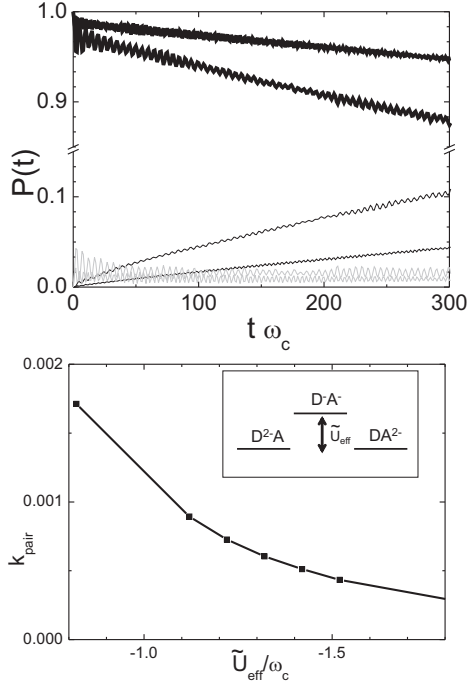}}
 \caption{ Upper panel:
Low-temperature population probabilities $P(t)=$ $d_D (t)$ (thick
black line), $d_A(t)$
 (thin black line) and $n_{DA}(t)$ (gray line) as functions of time.
$\tilde{U}=-0.9 \omega_c $ and $-1.5 \omega_c$ from bottom to top
for $d_D$ as well as from top to bottom for
 $d_A$.
Lower panel:
 Electron pair rate $k_{\rm pair}$ (for the transfer from
$D^{2-}A \rightarrow DA^{2-}$) as a function of $\tilde{U}_{\rm
eff}$. The parameters for both panels are $\alpha=0.16$, $T\approx 3
\cdot 10^{-8} \omega_c$, $\Delta=0.1\omega_c$ and $\varepsilon=0$.
 Inset: Energy levels of states $D^{2-}A$, $D^-A^-$ and $DA^{2-}$.
  } \label{fig_rate3}
\end{figure}

\begin{figure}
\vspace*{0cm} \hspace*{0cm} \epsfxsize=7cm
\centerline{\epsffile{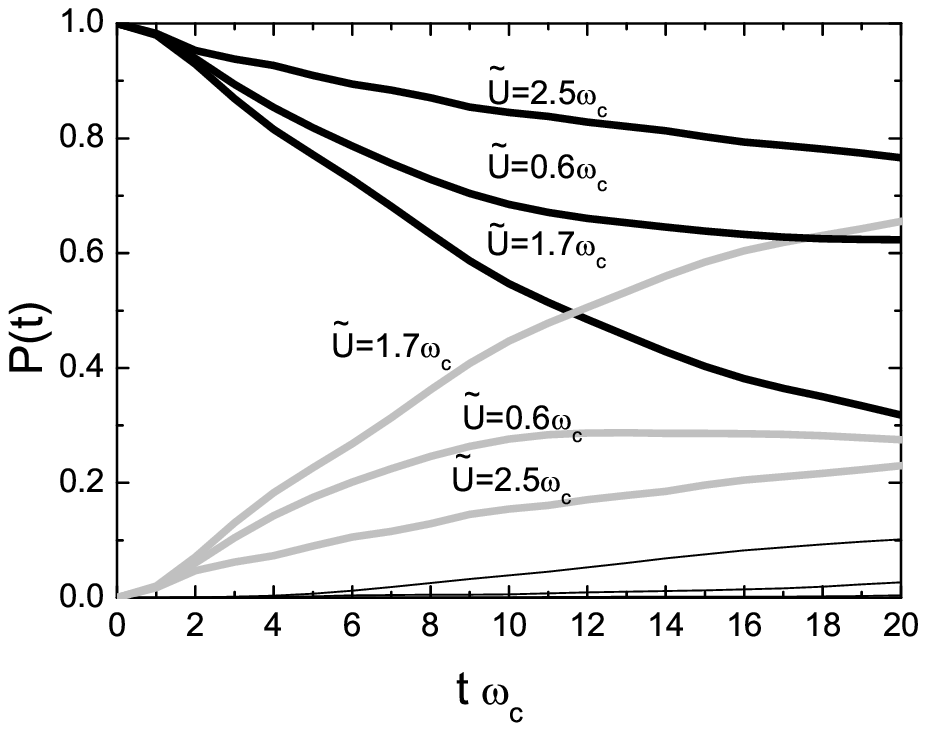}}
\centerline{\epsffile{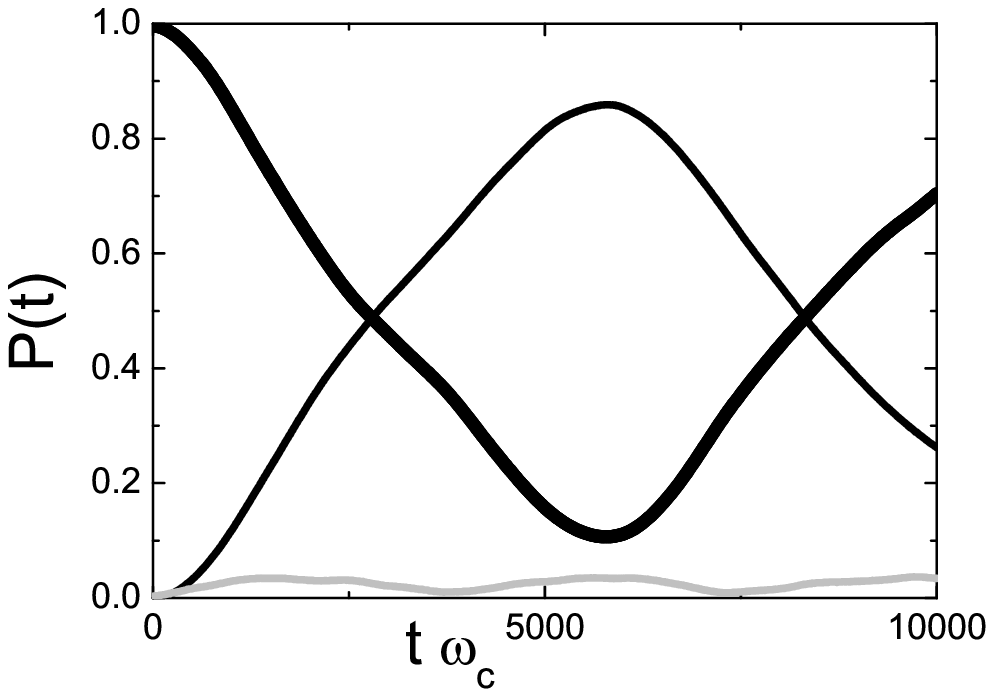}}
\centerline{\epsffile{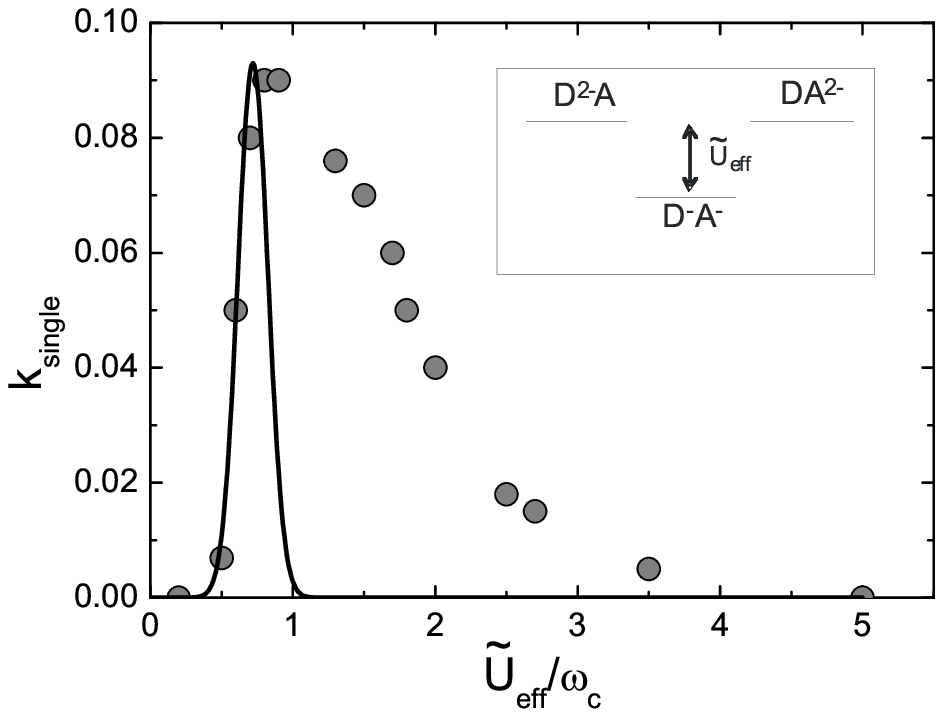}}
 \caption{Upper panel:
Low-temperature population probabilities $P(t)=$ $d_D (t)$ (thick
black line), $d_A(t)$
 (thin black line) and $n_{DA}(t)$ (gray line) as functions of time.
For $d_A$ from top to bottom $\tilde{U}=0.6 \omega_c $, $1.7
\omega_c$ and $2.5
 \omega_c$.  The other parameters are $\alpha=0.36, T\approx 3 \cdot
10^{-8}\omega_c$, $\Delta=0.1\omega_c$ and $\varepsilon=0$.
 Middle panel: Low-temperature population
probabilities with  $U=5\omega_c$. Lower panel: Single-electron rate
$k_{\rm single}$ for the transfer from $D^{2-}A \rightarrow D^-A^-$
 deduced by fitting the population probabilities with the
kinetic equations eq. (\ref{kinetic}) (squares) and Marcus rate eq.
(\ref{Marcus1}) (full line) with $T=0.008 \omega_c$ as a function of
the on-site Coulomb repulsion. The Marcus rate is normalized so that
both curves have the same maximal rate.  Inset: Energy levels of
states $D^{2-}A$, $D^-A^-$ and $DA^{2-}$.} \label{fig_rate}
\end{figure}
Next we study the effect of changing $\tilde{U}$ at constant system
bath coupling $\alpha=0.04$ (Fig.~\ref{fig_a01}), $\alpha=0.16$
(Fig.~\ref{fig_rate3}) and $\alpha=0.36$ (Fig.~\ref{fig_rate}) and
$\Delta=0.1 \omega_c$.

In the lower damping case (Fig.~\ref{fig_a01}), the transfer is
reflected by damped electron pair oscillations for
$\tilde{U}=-\omega_c$ in Fig.~\ref{fig_a01} (a). Increasing
$\tilde{U}=-0.5 \omega_c$ in Fig.~\ref{fig_a01} (b) leads to an
increase of the population probability of $D^-A^-$ and to a change
of the fast oscillations whit an approximate frequency of
$\tilde{U}_{\rm eff}$. When $\tilde{U}_{\rm}$ becomes positive
$\tilde{U}=0.5
 \omega_c$ (Fig.~\ref{fig_a01} (c)) the single-electron transfer
becomes fast and the main process unless $\tilde{U}_{\rm eff}$ is
not too large. In fact at $\tilde{U}=\omega_c$ the rate from
$D^{2-}A$ to $D^-A^-$ becomes smaller (Fig.~\ref{fig_a01} (d)) and
additional electron pair transfer is observed. The graphs
Figs.~\ref{fig_a01}(a) and (d) can be understood in terms of
Eq.~(\ref{occ_tU}) since $|\Delta/\tilde U|\ll 1$ and $\alpha=0.04$
is small. By the weak coupling to the environment, $\Delta$ is
slightly reduced, and the oscillation amplitude decays
exponentially. The difference between the two panels (a) and (d)
arises from (i) $\tilde{U}_{\rm eff}=\tilde U -2\alpha \omega_c$
instead of the $|\tilde U|$  entering Eq.~(\ref{occ_tU}) and (ii)
from the dissipation which favors the relaxation into the new
thermodynamic ground state: while the oscillation frequencies are
roughly the same for $|\tilde U|=\omega_c$, the delocalized states
have a lower energy in Fig.~\ref{fig_a01}(d) so that $n_{DA}(t)$ has
to increase to its new equilibrium value. The approximations made in
Eq.~(\ref{occ_tU}) do not hold any longer for the parameters in
Figs.~\ref{fig_a01}(b) and \ref{fig_a01}(c). The electronic dynamics
is governed by additional frequencies and becomes more complex.
However, the results can still be analyzed and understood within the
analytical results of $d_D(t)$, $d_A(t)$ and $n_{DA}(t)$ for $\alpha
\rightarrow 0$.

When the coupling $\alpha$ is increased to $\alpha=0.16$, a
different picture emerges. Very high frequency oscillations with a
small amplitude are superimposed on a slowly decaying $d_{D}(t)$
depicted in in the upper panel Fig. \ref{fig_rate3}. Averaging over
those oscillations, we can fit the population probabilities to the
kinetic equations (\ref{kinetic}). By this procedure, we extract the
phenomenological rates as function of $\tilde{U}_{\rm eff}$ for
fixed $\alpha=0.16$. As shown in the lower panel of Fig.
\ref{fig_rate3} the concerted transfer rate  $k_{[D^{2-}A
\rightarrow DA^{2-}]}^{\rm pair}$ increases with increasing
$\tilde{U}_{\rm eff}$ ($\tilde{U}_{\rm eff}<0$). This was expected
from the rate equation (\ref{marcussuper}) in the classical limit.

The transfer is found to be incoherent and sequential in the higher
damped case $\alpha=0.36$ for not too large $\tilde{U}$. The
population probabilities are shown for $\tilde{U}=0.6 \omega_c, 1.7
\omega_c$ and $2.5 \omega_c$ in the upper panel of
Fig.~\ref{fig_rate}. By fitting the curves with the help of the
kinetic equations, eq.~(\ref{kinetic}), we obtain the rate of the
single-electron transfer $D^{2-}A$ to $D^-A^-$ which is a
non-monotonic function of $\tilde{U}_{\rm eff}$ with a maximum at
$\tilde{U}_{\rm eff}=E_{\alpha 1}\approx 0.72 \omega_c$ (see lower
panel). It is plotted together with the Marcus rate at
$T=0.008\omega_c$ (For varying temperatures we found that the fitted
rate is approximately constant for temperatures $T< 0.008 \omega_c$
in the considered parameter space.). Although the qualitative
behavior is captured by the Marcus rate the asymmetric shape of the
NRG result is more realistic in the nuclear tunneling regime. As
$\tilde{U}$ increases further the sequential transfer becomes
negligible in the inverted region. As a matter of fact, an
increasing value of $\tilde U$ shifts the system away from the phase
transition line deeper into the delocalized phase as can be seen in
the equilibrium phase diagram Fig.~\ref{phase}. Here, the dynamics
is dominated by coherent pair oscillations with a very small
frequency, displayed for $\tilde{U}=5\omega_c$ in the middle panel
of Fig.\ref{fig_rate}.
\begin{figure}
\vspace*{0cm} \hspace*{0cm} \epsfxsize=9cm
\centerline{\epsffile{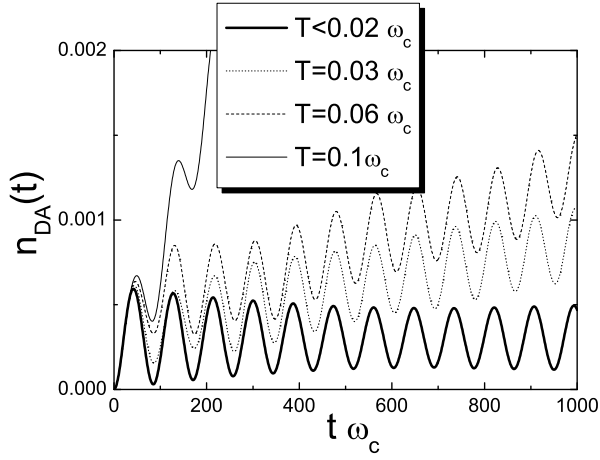}}
 \caption{Population probability $n_{DA}$ of the state
D$^-$A$^-$ as a function of time $t$ for temperatures between
$T<0.02\omega_c$ and $T=0.125\omega_c$. The parameters are
$\tilde{U}=-0.01\omega_c$, $\Delta=0.001\omega_c$, $\alpha=0.03$.}
\label{term}
\end{figure}

Finally, the effect of temperature is studied in Fig. \ref{term}
where $\tilde{U}=-0.01\omega_c$, $\Delta=0.001\omega_c$. The
temperature is varied from $3\cdot 10^{-8} \omega_c$ to $0.125
\omega_c$. For $T=3\cdot 10^{-8} \omega_c$ to $T\leq 0.02 \omega_c$
the population probability is temperature independent. As long as
$\tilde{U}_{\rm eff}>T$, pair transfer is observed (the probability
of $D^-A^-$ stays constant). As $T > \tilde{U}_{\rm eff}$ the states
$D^-A^-$ are seen to contribute and are thermally populated.
\begin{table}
\begin{tabular}{lll}
\hline \hline  $|\tilde{U}_{\rm eff}|\leq \Delta,$ &
single-electron transfer  \\
 \hline
$T>\tilde{U}_{\rm eff}$   &single-electron transfer\\
\hline
   \underline{$\tilde{U}_{\rm eff}>0$:}  \\
 $ \tilde{U}_{\rm eff}<E_{\rm \alpha 1}$ &  single-electron transfer \\
 & $k_{[D^{2-}A \rightarrow D^-A^-]}$ faster and \\&$k_{[D^{-}A^- \rightarrow D^-A^{2-}]}$ slower \\& with increasing $\tilde{U}_{\rm eff}$\\
 $ \tilde{U}_{\rm eff}>E_{\rm \alpha 1}$ &  single-electron transfer
\\  & $k_{[D^{2-}A \rightarrow D^-A^-]}$ \\ & and $k_{[D^{-}A^- \rightarrow D^-A^{2-}]}$ \\& slower with increasing $\tilde{U}_{\rm eff}$     \\
   $\tilde{U}_{\rm eff}\gg \Delta, |\tilde{U}_{\rm eff}| \gg E_{\alpha 1}$,\\ $T< |\tilde{U}_{\rm eff}|$
&  electron-pair transfer \\
 & (in addition slow \\ & single-electron transfer)
 \\
\hline
  \underline{ $\tilde{U}_{\rm eff}<0$:}  \\
 $ |\tilde{U}_{\rm eff}|<E_{\rm \alpha 1}$ &  single-electron transfer  \\
 & $k_{[D^{-}A^- \rightarrow D^-A^{2-}]}$ faster and \\&$k_{[D^{2-}A^- \rightarrow D^-A^{-}]}$ slower \\& with increasing $|\tilde{U}_{\rm eff}|$\\
 $ |\tilde{U}_{\rm eff}|>E_{\rm \alpha 1}$ &  single-electron transfer
\\  & $k_{[D^{-}A^- \rightarrow D^-A^{2-}]}$ \\ & and $k_{[D^{2-}A^- \rightarrow D^-A^{-}]}$ \\ & slower with increasing $|\tilde{U}_{\rm eff}|$     \\
 $|\tilde{U}_{\rm eff}|\gg \Delta,  |\tilde{U}_{\rm eff}| \gg E_{\alpha 1}$, \\ $ T< |\tilde{U}_{\rm eff}|$ &  electron-pair transfer \\

 \hline \hline
\end{tabular}
\caption{Summary of the results. The effective Coulomb repulsion is
defined by $\tilde{U}_{\rm eff}=U-V-2\alpha \omega_c$. The
corresponding reorganization energy is $E_{\alpha 1}=2 \alpha
\omega_c$ the bias is $\varepsilon=0$. Starting with two electrons
on the donor the system performs either a sequential single-electron
transfer ($D^{2-}A\rightarrow D^-A^- \rightarrow DA^{2-} $) or a
pair transfer ($D^{2-}A\rightarrow DA^{2-}$) depending on
$\tilde{U}_{\rm eff}$.}
\end{table}

\section{Summary and Conclusion \label{sum}}

In this paper, we have studied the electron transfer properties of
two excess electrons in a redox system modeled as a dissipative
two-site Hubbard model -- a model which can be viewed as the
simplest generalization of the spin-boson model to include
many-particle effects. These many-particle effects are due to
on-site and inter-site Coulomb interactions, $U$ and $V$
respectively, as well as the effective interactions  induced by the
coupling to a common bosonic bath. These interaction parameters can
be calculated by {\it ab initio} methods for a specific system (see,
for example, \cite{Starikov,Fulde}). In our two-site model only the
difference $\tilde{U}=U-V$ enters the dynamics. In the presence of a
bosonic bath, the effective energy $\tilde{U}$ is renormalized to
$\tilde{U}_{\rm eff} = \tilde{U}-2\alpha \omega_c$. An effective
attractive interaction $\tilde{U}_{\rm eff}<0$ favors the
localization of two electrons on the same site, a repulsive
$\tilde{U}_{\rm eff}>0$ favors the distribution of electrons on
different sites.

The intricate correlated dynamics of two electrons depends on the
activation energy. This is because, for the tunneling of an electron
between two states, energy fluctuations are necessary for the
reorganization of the donor-acceptor system and is influenced by an
energy difference between the states. Therefore, the transfer
characteristics in the unbiased case depends strongly on the
effective on-site Coulomb repulsion $\tilde{U}_{\rm eff}$. Three
rates have to be considered: the forward and backward rate between
the double occupied states ($D^{2-}A$, $DA^{2-}$) and the two
intermediate degenerate states ($D^-A^-$) as well as the direct rate
between $D^{2-}A$ and $DA^{2-}$. How these rates depend of
$\tilde{U}_{\rm eff}$ is summarized in Table I.

We have performed calculations for the probabilities $P(t)$ of
doubly and singly occupied donor and acceptor states using the
time-dependent Numerical Renormalization Group method
\cite{AS1,AS2}. This information helps us to identify conditions
under which the systems performs (a) concerted two-electron
transfer, (b) uncorrelated, sequential single-electron transfer or
(c) fast concerted two electron followed by a single-electron
transfer. With the time-dependent NRG method we can describe the
crossover from damped coherent oscillations to incoherent relaxation
as well as to localization (at $T\rightarrow 0$). The temperatures
are chosen to be $0.1 \omega_c >T>3\cdot 10^{-8} \omega_c $. For
larger temperatures, when the bosonic bath can be treated
classically, the Marcus rates are applicable.

For $\tilde{U}_{\rm eff}\gg \Delta, E_{\alpha 1},T$ concerted
electron transfer occurs in both methods: in the nuclear tunneling
regime within the NRG as  well as in the limit of a classical bath
within the Marcus theory. As long as $T< \tilde{U}_{\rm eff}$,
however, thermal activation is absent and nuclear tunneling is the
main process. Only a full quantum mechanical calculation yields the
correct relaxation rates which are governed by quantum-fluctuation,
dephasing and energy exchange with the environment.

For small $\Delta/|\tilde{U}_{\rm eff}|$ we found an effective pair
hopping via virtual population of the low lying or high lying states
$D^-A^-$. When the equilibrium probability for the states $D^-A^-$
is finite, a slow single-electron accompanies the faster pair
transfer. In contrast to the single-electron transfer with a
frequency of the order $\Delta$, the frequency of the pair transfer
is of the order $4\Delta^2/|\tilde{U}_{\rm eff}|$.

The concerted transfer becomes more uncorrelated and sequential at
short times at high  temperatures  ($T>\tilde{U}_{\rm eff}$),
increasing coupling to the bosonic bath ($E_{\alpha 1}\geq
 \tilde{U}_{\rm eff}$) or larger single-electron hopping ($\Delta
\geq \tilde{U}_{\rm eff}$). 
The sequential transfer  rate is  non-monotonic with increasing
$\tilde{U}_{\rm eff}$. At first, the transition rate  from $D^{2-}A$
to  the delocalized states $D^-A^-$ increases for small
$\tilde{U}_{\rm eff}>0$, reaches a maximum for $\tilde{U}_{\rm
  eff}=E_{\alpha 1}$ before it decreases again. The rate for the
consecutive process $D^-A^-\rightarrow DA^{2-}$, however,  decreases
with increasing $\tilde{U}$. For a negative effective Coulomb matrix
element $\tilde{U}_{\rm eff}$, the transfer rate of the second
process $D^-A^-\rightarrow DA^{2-}$ is maximal for $\tilde{U}_{\rm
eff}=-E_{\alpha 1}$. In this parameter regime we expect that the
second electron follows very shortly after the first electron was
transferred.

The transfer kinetics of more than two excess charges in, for
example, biochemical reaction schemes or molecular electronics
applications is controlled by the molecule specific Coulomb
interaction and its polar environment. Our study reveals the
conditions for concerted two-electron transfer and sequential
single-electron transfer. Concerted two-electron transfer is
expected in compounds where the difference of the inter-site Coulomb
repulsion and effective on-site repulsion are much larger than the
single-electron hopping and larger than the temperature and
reorganization energy. Furthermore, we have shown that non-monotonic
characteristics of sequential single-electron transfer strongly
depends on the Coulomb interaction. A further study will include the
influence of a finite energy difference $\varepsilon$ between the
donor and acceptor site. 
We will also report on the influence of Coulomb repulsion and
many-particle effects on the long-range charge transfer using a
longer Hubbard chain as bridge between donor and acceptor centers.

\section*{Acknowledgments}
S. T. is grateful to the School of Chemistry of the Tel Aviv
University and to the Racah Institute of Physics of the Hebrew
University of Jerusalem for the kind hospitality during her stay and
partial support (Tel Aviv University). This research was supported
by the German science foundation (DFG) through SFB 484 (S.~T.,
R.~B.), AN 275/5-1 and 275/6-1 (F.~B.~A.), by the National Science
Foundation under Grant No.~NFS PHYS05-51164 (F.~B.~A.), by the
German-Israel foundation (A.~N.), the Israel science foundation
(A.~N.) and the US-Israel Binational Science foundation (A.~N.).
F.~B.~A. acknowledges supercomputer support by the NIC,
Forschungszentrum J\"ulich under project no.\ HHB000. We acknowledge
helpful discussions with A.~Schiller and D.~Vollhardt.


\end{document}